\documentclass{aa}

\usepackage{graphicx}
\usepackage{txfonts}
\usepackage{lipsum}
\usepackage{subcaption}         
\usepackage{lscape}
\usepackage{placeins}

\usepackage[colorlinks]{hyperref}
\hypersetup{
  colorlinks   = true, 
  urlcolor     = blue, 
  linkcolor    = blue, 
  citecolor   = cyan 
}

\usepackage{xcolor}
\usepackage{comment}
\usepackage{keyval}
\usepackage{natbib}

\newcommand{\pc}{\,\mathrm{pc}}
\newcommand{\kpc}{\,\mathrm{kpc}}
\newcommand{\Mpc}{\,\mathrm{Mpc}}

\newcommand{\Gyr}{\,\mathrm{Gyr}}

\newcommand{\Msun}{\, M_{\sun}}

\newcommand{\nc}{\mbox{\sc \small NewCluster}}

\begin{document}
   \title{Introducing NewCluster: the first half of the history of a high-resolution cluster simulation}

   \author{San Han\inst{1}\fnmsep\thanks{Email address: san.han@iap.fr}
        \and Sukyoung K. Yi\inst{2}\fnmsep\thanks{Email address: yi@yonsei.ac.kr}
        \and Yohan Dubois\inst{1}
        \and Jinsu Rhee\inst{1,3}
        \and Seyoung Jeon\inst{2}
        \and J. K. Jang\inst{2}
        \and Gyeong-Hwan Byun\inst{2}
        \and Corentin Cadiou\inst{1}
        \and Juhan Kim\inst{4}
        \and Taysun Kimm\inst{2}
        \and Christophe Pichon\inst{1,5}
        }

   \institute{Institut d’Astrophysique de Paris, Sorbonne Université, CNRS, UMR 7095, 98 bis bd Arago, 75014 Paris, France
            \and Department of Astronomy and Yonsei University Observatory, Yonsei University, Seoul 03722, Republic of Korea
            \and Korea Astronomy and Space Science Institute, 776, Daedeokdae-ro, Yuseong-gu, Daejeon 34055, Republic of Korea
            \and Korea Institute of Advanced Studies (KIAS) 85 Hoegiro, Dongdaemun-gu, Seoul 02455, Republic of Korea
            \and Kyung Hee University, Dept. of Astronomy 
            \& Space Science, Yongin-shi, Gyeonggi-do 17104, Republic of Korea\\ }

   \date{Revised version November 23, 2025}

   \abstract
    {}
   {We introduce \nc, a new high-resolution cluster simulation designed to serve as the massive halo counterpart of the modern cosmological galaxy evolution framework.}
   {The zoom-in simulation targets a volume of $4.1\sigma$ overdensity region, which is expected to evolve into a galaxy cluster with a virial mass of $5 \times 10^{14} M_\odot$, comparable to that of the Virgo Cluster.
   The zoom-in volume extends out to 3.5 virial radii from the central halo.
   The novelties of \nc\ are found in its resolutions. Its stellar mass resolution of $2 \times 10^{4} M_\odot$ is effective for tracing the early assembly of massive galaxies as well as the formation of dwarf galaxies.
   The spatial resolution of 68\,parsecs in the best-resolved regions in the adaptive-mesh-refinement approach is powerful to study the detailed kinematic structure of galaxies.
   The time interval between snapshots is also exceptionally short—15 Myr—which is ideal for monitoring changes in the physical properties of galaxies, particularly during their orbital motion within a larger halo.
   The simulation has up-to-date feedback schemes for supernovae and active galactic nuclei. The chemical evolution is calculated for ten elements, along with dust calculation that includes the formation, size change, and destruction.
   To overcome the limitations of the Eulerian approach used for gas dynamics in this study, we employ Monte Carlo-based tracer particles in \nc, enabling a wide range of scientific investigations.}
   {The simulation has passed $z=0.8$, well over half of the cosmic history. We release the early data with the expectation that they will facilitate studies of the early evolution of galaxies and overdensities.}
   {}


   \maketitle
\nolinenumbers
\section{Introduction}\label{sec:intro}

Numerical simulations have become an indispensable tool in modern astronomy for exploring the complex and non-linear nature of the Universe.
A pioneering study by \cite{TT72}, which modeled the morphologies of interacting galaxies, demonstrated the potential of numerical simulations to capture such non-linearity.
Since then, continuous advancements in computational methods have established numerical methods as an important complement to observational and theoretical approaches \citep[e.g.,][]{WR78, Davis85,Hernquist87,Cole91,WF91,Kauffmann93}.

These early successes have laid the foundation for the advancement of contemporary cosmological simulations.
The cosmological principle assumes that both the spatial distribution of matter and its power spectrum approach homogeneity on sufficiently large scales.
This justifies the use of a large numerical volume to approximate a representative subregion of the Universe.
Provided this, cosmological simulations are initialised with primordial density fluctuations consistent with the early Universe and evolved forward in time using the governing equations of gravity and hydrodynamics, coupled with astrophysical subgrid models.
These calculations yield the formation and evolution of self-consistent large-scale structure within the simulated volume \citep[e.g.,][]{Peebles82, Davis85,White87}.

Observational validation of the $\Lambda$CDM cosmology, including precise measurements of cosmological parameters \citep[e.g.,][]{2011ApJS..192...18K}, significantly accelerated the development of early cosmological numerical simulations.
A prominent example is the \textsc{Millennium} simulation \citep[][]{Springel05}, based on the $\Lambda$CDM model, which successfully reproduced large-scale structures consistent with that observed in the survey \citep[][]{Springel06}.
This achievement sets a stage for practically attainable theoretical investigation of nonlinear cosmological structures.
Although the simulation itself included only collisionless dark matter (DM) particles, subsequent studies incorporated semi-analytic models to populate galaxies, demonstrating an important link between the non-linear nature of the Universe and galaxy evolution \citep[e.g.,][]{Bower06, deLucia07}.

The next generation of cosmological simulations incorporated hydrodynamics to model baryonic physics, including cooling and heating, star formation, and feedback, through dedicated astrophysical prescriptions.
Seminal large-volume simulations that pioneered this approach include \textsc{Eagle} \citep[][]{Schaye15}, \textsc{Horizon-AGN} \citep[][]{2014MNRAS.444.1453D}, and \textsc{Illustris} \citep[][]{Vogelsberger14}.
These simulations typically achieved spatial resolutions of $\sim1\,{\rm kpc}$, sufficient to capture the global properties and statistical trends of galaxy populations in cosmological contexts \citep[][]{Vogelsberger14b, Furlong15, Dubois16}.
The larger volumes further allowed for systematic studies of environmental effects on galaxy evolution by treating environment as a controllable parameter \citep[e.g.,][]{2022ApJ...941....5J}.
However, a resolution of $\sim 1\,{\rm kpc}$ remains insufficient to resolve the internal structure and processes with sub-kpc physical scales within galaxies.
For example, in a Milky Way-like galaxy with a size of $\sim 10\,{\rm kpc}$, only a few resolution elements are available to represent its internal structure.

Recent cosmological simulations have reached the sub-{\rm kpc} resolution \citep[e.g.,][]{Tremmel17, Pillepich19}.
With this improvement in resolution, galactic substructures begin to be resolved in a cosmological context \citep[e.g.,][]{Du21}, enabling detailed analyses of the kinematic and spatial properties of galactic components.
These sub-{\rm kpc} simulations have also resolved structures beyond galaxies themselves, such as circumgalactic medium (CGM) \citep[e.g.,][]{Peeples19}, making it feasible to study the baryon cycle of galaxies.
However, despite these advances, it is sometimes insufficient to resolve finer galactic structures, such as the multiphase nature of interstellar medium (ISM) or the thin disc of spiral galaxies.

Subsequent cosmological hydrodynamic simulations have pushed the resolution below $\sim 100\,{\rm pc}$ \citep[e.g.,][]{Wang15, Wetzel16, 2021A&A...651A.109D, Yi24}, with some achieving resolutions as fine as sub-10\,pc \citep[][]{Hopkins18b}.
This level of spatial resolution enables resolving both the distinction of thin and thick discs of galaxies \citep[e.g.,][]{Park21, renaud2021, Yi24} or compact systems \citep[][]{Jang2024} and the multiphase structure of galactic gas, including the warm and cold ISM \citep{2021A&A...651A.109D}.
Such high resolution has facilitated the implementation of more advanced subgrid models.
Mechanical supernova feedback, which has been reported to be more effective in suppressing star formation in low-mass galaxies, has been implemented in several high-resolution simulations, enabled by the ability to assign momentum to gas elements in the immediate vicinity of star particles \citep{2014ApJ...788..121K,Hopkins18a}.
Direct measurements of the differential velocity within the ISM have been used to implement the gravo-thermo-turbulent model of star formation efficiency \citep{2017MNRAS.466.4826K,Kretschmer20}, allowing detailed investigations on star formation activity within galaxies \citep[][]{Kraljic24}.

Furthermore, the corresponding increase in baryonic (and DM) mass resolution is also important, as poor resolutions introduce numerical biases \citep[e.g.,][]{Klypin99,Ludlow19}.
Fine mass resolutions allow for the detection of structures with extremely low surface brightness, such as intracluster light (ICL), ultra-diffuse galaxies (UDGs), tidal streams, and ultra-faint dwarfs.
A stellar particle mass of $\sim10^4\,\Msun$ enables the simulation to reach surface brightness levels as faint as $34\,{\rm mag\,arcsec^{-2}}$ at $z=0.1$.
This is enough to reproduce results from deep imaging surveys such as Large Synoptic Survey Telescope (LSST), Euclid, and James Webb Space Telescope (JWST) \citep{2022A&A...657A..92E}.

Therefore, more realistic and advanced cosmological simulations at the current stage require a sub-100\,pc resolution and a high mass resolution.
A major challenge arising from these requests is the high computational cost due to the enormous number of elements and associated calculations.
A good bypass to satisfy the needs above and to avoid this difficulty is running a simulation for a small interested region with high resolutions, while the rest of the volume is treated with relatively poorer resolutions: that is, a zoom-in approach \citep[e.g.,][]{Navarro94, Barnes17, CY17, Hopkins18b, Tremmel19, 2021A&A...651A.109D}.

The majority of recent high-resolution (sub-100\,pc) zoom-in simulations focus on individual galaxies or average-density fields of the Universe.
This has resulted in a lack of representative high-resolution counterparts for dense environments such as galaxy clusters.
Galaxy clusters are thought to have formed from the growth of primordial over-density in the early Universe \citep{2012ARA&A..50..353K}, implying that developing clusters were once among the most active sites of star formation \citep{2022ApJ...941....5J}.
Therefore, proto-clusters are the primary candidates for the sites for containing high-redshift galaxies that are frequently detected by the JWST \citep{Labbe23, Carniani24}.

In this regard, we carry out the \nc\ simulation, a high-resolution cosmological hydrodynamic simulation designed to reproduce the formation and evolution of a galaxy cluster down to near-local redshift.
\nc\ serves as a massive halo counterpart to existing high-resolution zoom-in simulations on average-density fields, \textsc{NewHorizon} and \textsc{NewHorizon2}, particularly because it shares many of its subgrid prescriptions with them. 
The simulation aims to monitor the formation and evolution of a massive cluster and galaxy evolution inside the extreme environment.

We expect morphological transformations of galaxies in clusters, which are responsible for the prevalence of early-type galaxies with old stellar populations \citep{Dressler1980, Lewis02, Hogg04, Kauffmann04}.
This transition is thought to be triggered by the interplay of both internal and external processes.
The energy and outflows released by stellar activity and central supermassive black holes are considered major factors in the quenching of star formation \citep{2005ApJ...618..569M,2011MNRAS.414..195T}.
The released gas forms the hot intracluster medium (ICM), which in turn drives an external quenching process for cluster members.
Moving at very fast orbital velocity ($\sim10^3\,{\rm km\,s^{-1}}$), galaxies experience rapid gas stripping through hydrodynamic ram pressure and DM and stellar stripping through gravitational tidal interactions \citep{Smith16, Rhee17, Han18, Jung18}.
Such processes are likely also responsible for jellyfish \citep{Chung09} and perhaps ultra-diffuse galaxies \citep{vanDokkum2015}.
To closely trace the gravitational and hydrodynamic interactions that impact galaxy evolution, it is essential to achieve sufficient spatial resolution to resolve the kinematics and structural properties of galaxies.
Such a level of detail was previously unavailable in simulations of galaxy clusters.

The impact of the cluster merger (starting at $z=0.8$) is also open to wild speculations.
While \nc\ lacks the capability for statistical analysis, its high spatial resolution offers the opportunity to examine the merging process in greater detail, particularly regarding how member galaxies are affected by the global redistribution of the gravitational mass and ICM.

In this paper, we summarise the results of the \nc\ simulation up to $z = 0.8$, which corresponds to the half of the age of the universe and represents the latest available snapshot.
At this redshift, a halo with a mass comparable to that of a galaxy cluster is present, which enables the simulation to reproduce the formation of an observed galaxy cluster analogue and its associated phenomena.
This paper is organised as follows.  
In Sect.~\ref{sec:method}, we describe the simulation code, astrophysical prescriptions, and post-processing methods.  
In Sect.~\ref{sec:results}, we present a general overview of the simulation and the galaxy scaling relations at $z=0.8$, along with key examples of applications of the cluster simulation to various topics.
In Sect.~\ref{sec:summary}, we summarise the main findings of the paper.

\section{Methods}\label{sec:method}
\subsection{Code: RAMSES-yOMP}
RAMSES \citep{2002A&A...385..337T} is an astrophysical simulation code based on the adaptive mesh refinement (AMR) technique.
In this study, we use a specialised branch of the code, called RAMSES-yOMP \citep{yomp}, which has been extensively modified to enhance simulation performance.
The improvement is primarily due to the implementation of a hybrid parallelization scheme that utilises a dual-level multiprocessing structure to enhance the parallel scalability, combining the message passing interface (MPI) and open multi-processing (OMP) libraries.
A series of modifications to yOMP are described in \cite{yomp}, including updates on subroutines such as the load-balancing scheme and the conjugate gradient Poisson solver, aimed at improving the overall performance of the simulation. 
Load balancing is enhanced by assigning greater computational weights to particles that require more processing, particularly those involved in complex feedback and density estimation.
In a test suite simulating a cosmological volume with 1,536 computing cores, the modified code achieved approximately a two times speed-up compared to the pre-modified version.

\subsection{Initial condition and zoom-in region}
The initial condition is generated using MUSIC \citep{2011MNRAS.415.2101H} at $z=50$ with the WMAP7 cosmology \citep{2011ApJS..192...18K} with parameters $\Omega_\Lambda=0.728$, $\Omega_m=0.272$, $\sigma_8=0.810$, $n_s=0.967$, $h_0=0.703$ and $\Omega_b=0.0455$.
The transfer function from \cite{1998ApJ...496..605E} and CAMB \citep{2000ApJ...538..473L} is used to derive the initial distribution of DM and baryons, respectively.
The zoom-in technique is utilised to efficiently simulate cluster regions in high resolution.
First, we ran DM-only simulations of a box with a volume of $(100\,h^{-1}\Mpc)^3$ with periodic boundary conditions.
We use the Rockstar halo finder \citep{2013ApJ...762..109B} to detect halos within the volume at $z=0$.
Since the simulation is performed on only one galaxy cluster, we ran 15 DM-only simulations in total, and selected one target halo based on visual inspection.
The target cluster whose evolution matches the following criteria was selected:
1. A cluster in a relaxed state at z = 0, without ongoing major mergers.
2. A cluster that had experienced a major merger during its assembly phase.
A spherical region centred on the density peak of the target cluster is set with a radius of $3.5R_{\rm vir}$, where $R_{\rm vir}$ refers to the virial radius\footnote{In Rockstar, the virial radius is defined the range within which contains $\Delta_{\rm vir}(z)\rho_c$ \citep{Bryan1998, 2013ApJ...762..109B}.} of the target cluster, to record the identity of the confined particles and trace their initial distribution at $z=50$.
The zoom-in region, corresponding to a $4.1\sigma$ overdensity above the Universal mean, is defined as a convex hull generated based on the distribution of the recorded particles in the initial volume.
The selected convex hull roughly covers a comoving volume of $23,300\Mpc^3$ with a total mass of $8.6\times10^{14}\Msun$.

The volume within the zoom-in region is initialised with grids and particles with a base level of 12, the scale that corresponds to the side length of the box divided into $2^{12}$ cells.
This corresponds to an initial spacing of $34.7\, {\rm kpc}$ in comoving units.
Padding grids are deployed around the finest zoom-in region to provide a gradual transition in resolution.
This results in a continuously decreasing level of refinement in grids and an increasing mass of DM particles, from the finest level of 12 at the centre to coarser levels further away from the zoom-in region, reaching the minimum level of 8.

A single DM particle is placed in each cell with a slight offset from the centre to account for the density distribution of the initial condition.
At level 12, DM particles have a mass of $\sim1.3\times10^6\Msun$, and increase by a factor of eight as their initially placed refinement level decreases by one.
As the simulation progresses, DM particles with lower refinement levels may reach inside the zoom-in region, which may impact the galaxies near the outskirts of the zoom-in region, leading to potential inconsistencies in the computation of gravitational potential and refinement.
To address this, we impose our own criteria for the fraction of low-level DM particles (i.e., contamination) to exclude affected galaxies (see Sect.~\ref{section:halogaldetection}).

We also assign an extra passive scalar variable to the hydro information in cells, indicating a refinement mask which is set to 1 within the zoom-in region and 0 elsewhere.
Grids with refinement masks greater than 0.05 are considered to be within the zoom-in region and become targets for refinement to higher resolution, as well as the application of baryonic physics such as star formation and active galactic nucleus (AGN) feedback.

The best spatial resolution, i.e.~the minimum cell size within the zoom-in region, depends on the scale factor of the simulation.
This is because the simulation grid size is built based on the comoving units.
To maintain physical resolution, we allow an additional level of refinement for cells each time the scale factor doubles, that is, at the transition epochs $a = 0.1$, $0.2$, $0.4$, and $0.8$, effectively unlocking finer spatial resolution at these epochs.
This results in a slightly different cell size depending on the current scale factor of the simulation, corresponding to a physical size of $53$--$107\, \pc$.
At $z=0$, the simulation achieves a final resolution of $68\,\pc$, corresponding to a grid level of 21.

The sudden increase in resolution at the specific scale factors may introduce numerical artifacts in the gas properties of galaxies.
To mitigate this, we smooth out the transitions at which refinement is applied.
We follow the approach of \cite{2018MNRAS.477..983S}, but additionally apply the same method to the Jeans length criterion.
During the transition phase, we continuously varied the parameter $N_{\rm jeans}$ from 0 to 1 and $m_{\rm refine}$ from 100 to 8, at the transition phase to perform smooth refinement, i.e.,
\begin{equation}
    C = C_{L} + \frac{1}{2}\left[1+\sin\left(\frac{(a - a_{0})\pi}{2w}\right)\right]\left(C_R-C_L\right),
\end{equation}
where $a$ is the current scale factor, $a_0$ is the centre of the transition phase (defined as 0.1, 0.2, 0.4, 0.8 at grid levels 18, 19, 20, 21), and $w$ is the half-width of the transition phase, set to $0.025$ in this work.
$C$ is the refinement threshold and can be either $N_{\rm Jeans}$ or $m_{\rm refine}$, where $C_L$ and $C_R$ denote the initial and final values for the refinement transition, respectively.
The parameter pair $[C_L, C_R]$ is set to $[0, 1]$ for $N_{\rm Jeans}$, and $[100, 8]$ for $m_{\rm refine}$.
$m_{\rm refine}$ defines the mass increase required for a cell to be refined.
For example, if $m_{\rm refine}=8$, the mass within a cell needs to be eight times higher than the initial cell mass to be eligible for refinement.
$N_{\rm Jeans}$ is the parameter for the Jeans length refinement criterion, in which the cell with a hydrogen gas density $n\ge 5\,\rm H\, cm^{-3}$ can be refined if
\begin{equation}
    l_{\rm Jeans} < N_{\rm Jeans}\Delta x \, ,
\end{equation}
where $l_{\rm Jeans}$ is the thermal Jeans length of the cell, and $\Delta x$ is the current size of the cell.
With this formula, cells with the shortest Jeans length, typically corresponding to the densest cells in a galaxy, are preferentially refined during the transition phase within a scale factor window of 0.05.

Since DM particles have considerably higher mass than baryonic components, the shot noise could introduce numerical artifacts in gravity calculations.
We set an upper limit on the contribution of DM density on the mesh at level 18, which corresponds to the comoving resolution of $0.54\,\kpc$.

\subsection{Astrophysical prescriptions}

\subsubsection{Stellar physics and chemical evolution}
The star formation rate (SFR) in the simulation is calculated based on the gravo-thermo-turbulent model of \cite{2017MNRAS.466.4826K} without applying the turbulent Jeans length criterion, thereby preserving scale invariance, as in \cite{2021A&A...651A.109D}.
The exact formation efficiency follows the equation provided by \cite{2012ApJ...761..156F} (The multi-ff PN model).
In each time step, the amount of stellar mass formed is checked within each gas cell above a star formation density threshold of $n_\star=5\,\rm H\,cm^{-3}$ to form a star particle.

Star particles are probabilistically formed according to a Poisson distribution.
The mean of the distribution is set by the SFR integrated by time, divided by the minimum mass, $2.49\times10^4\Msun$.
Consequently, the initial mass of each star particle is an integer multiple of this minimum mass.
The mass of a star particle may decrease over time due to mass ejection during its feedback processes, with up to $\sim40\%$ of its initial mass being ejected to the surrounding medium.

The stellar feedback in the simulation consists of two kinds: supernovae (SNe) and stellar winds.
The SN feedback is implemented as two independent types, Type II and Type Ia (as SNII and SNIa, respectively), both of which are implemented as the form of mechanical feedback scheme of \cite{2014ApJ...788..121K} that takes account of the empirical boost depending on the local density of the explosion \citep{2017MNRAS.466.4826K}.
Every star particle that has reached the age of 5\,Myr triggers the SNII explosions.
The produced feedback energy per mass is set to be $e_{\rm SNII}=0.045\times10^{51}\,{\rm erg}\Msun^{-1}$.

The frequency of SNIa is set to follow the delay time distribution (DTD) in the form of a power law \citep{Maoz12},
\begin{equation}
\Psi(t) = \phi \, t^\alpha,
\end{equation}
where the power index is fixed to be $\alpha=-1$ and the amplitude is chosen to be $\phi=2.35\times10^{-4}\Msun^{-1}$.
The value of DTD amplitude $\phi$ is derived as a result of the normalization of the lifetime frequency of $N_{\rm SNIa}/M_{\rm SSP}=0.0013\Msun^{-1}$ for a given simple stellar population of mass $M_{\rm SSP}$, assuming that it is integrated over the domain of $0.05-13.7\Gyr$.
The feedback energy of a single SNIa is set to $10^{51}\,{\rm erg}$.

For a realistic production of chemical elements within galaxies, we have included mass return from stellar ejecta of SNe type II and Ia, and stellar winds in the simulation.
Throughout the simulation, we assume a~\cite{2003PASP..115..763C} initial mass function for the integration of chemical yield from the stellar population.
Considering each stellar particle as a simple stellar population (SSP), chemical enrichment for ten species (H, D, He, C, N, O, Mg, Fe, Si, and S) is computed using the Starburst99 code \citep[][]{1999ApJS..123....3L, 2014ApJS..212...14L}.\footnote{D denotes deuterium that cannot be produced from ejecta of stellar origin, but only be present as an initial fraction of $({\rm D}/{\rm H}) = 2.6\times10^{-5}$ \citep{2016RvMP...88a5004C}. Helium is not explicitly traced in the simulation, but is considered as a residual component after the nine tracked elements, with an initial abundance of $Y=0.2477$ \citep{2007ApJ...666..636P}. For the other heavier elements, the gas is set to be metal-free at the initial state.}
Yields from stellar winds are derived assuming the Geneva stellar wind model \citep[][]{Schaller92, Maeder00}.
At the same time, energy produced from stellar wind is directly injected into the grid where the star is positioned.
SNII explosions also inject newly synthesised elements into surroundings based on the tabulated value for the SNII yields in \cite{2006ApJ...653.1145K}, assuming that each SSP has the progenitor mass range of $8-50\,M_{\odot}$, which corresponds to the ejecta mass fraction for each stellar particle is $M_{\rm ejecta}/M_{*} = 0.311$.
The yield from SNIa is derived from \cite{1999ApJS..125..439I} with the mass of the ejecta in every burst corresponding to the Chandrasekhar limit ($\sim1.4\Msun$), with corresponding decline of mass in each stellar particle.

\subsubsection{Dust formation and evolution}

The simulation also computes dust formation and evolution that evolves on-the-fly throughout cosmic time.
The dust model implemented for \nc\ is based on that described in \citet{Yohan24} (which we refer to for further technical details) and accounts for several key physical processes, including dust formation, growth, destruction, and size evolution.
Here, we provide a brief summary of the methodology.

Dust grain species are modeled with two distinct characteristic chemical compositions: carbonaceous and silicate grains.
We assume that silicate grains consist of a single form of amorphous olivine, with compound MgFeSiO$_4$, and that carbonaceous grains are composed purely of carbon.
To follow the grain size distribution of dust, the two different chemical compositions are further decomposed into two different grain sizes (adopting the two-size bin approximation of \citealp{Hirashita15}) of radius 5\,nm for small grains and 0.1\,$\mu$m for large grains. 

For dust grain production, we assume two channels: the formation from the stellar ejecta and growth through the accretion of gas-phase metals in the ISM.
We consider the dust production from SNe type II and Ia, and asymptotic giant branch (AGB) winds based on the feedback models and stellar yields described above with different dust condensation efficiencies that depend on the stellar origin and on the C/O ratio (see~\citealp{Yohan24} for the exact values). 
Accretion of gas-phase metals onto dust grains in dense and cold ISM regions is also implemented as a subgrid model.
Dust accretion rates are estimated based on the gas density, temperature, and metallicity in each gas cell and depend on the grain size and its material density ($s_{\rm gr}=3.3$ and $2.2\,\rm g\,cm^{-3}$ for silicate and carbonaceous grains respectively). 
The local gas velocity dispersion is used to sample a subgrid log-normal shape of the probability density function of the gas density, assuming isothermal turbulence.
We assume that the gas-phase accretion occurs only for (subgrid) gas densities lower than $n_{\rm thr} = 10^4\,{\rm H\,  cm^{-3}}$, considering the formation of ice mantles on grain surfaces, which inhibit metal accretion onto pre-existing grain cores \citep{Cuppen&Herbst07,Hollenbach09}.
The Coulomb enhancement factor, which accounts for the electrostatic effects acting differently on various grain species, is not included in the \nc\ dust model.
For dust destruction, we include thermal sputtering and supernova-driven destruction from both Ia and II explosions with increased efficiencies with decreasing grain sizes.
The thermal sputtering timescale with temperature is estimated following \citet{Tsai&Mathews95}.
SNe destroy the enclosed dust until the shock velocity decreases below the activation threshold of $100\,\rm km\,s^{-1}$, with identical destruction efficiencies for graphite and silicate grains.
Astration completely destroys the dust engulfed in the star formation process.

The two grain sizes of each grain chemical composition can exchange mass by shattering and coagulation, respectively, decreasing and increasing the grain sizes.
Shattering operates at intermediate gas densities ($n\lesssim10\,\rm H\,cm^{-3}$) where grain turbulent velocities ($\sigma_{\rm gr}\gtrsim10\,\rm km\,s^{-1}$) are sufficiently high to fragment the grains \citep{Jones1996,Yan04}.
In contrast, coagulation takes place in the dense gas phase ($n\gtrsim100\,\rm H\,cm^{-3}$) where turbulence is sufficiently low ($\sigma_{\rm gr}\lesssim1\,\rm km\,s^{-1}$) to preserve the grain structure and to allow grain size growth by collisions \citep{Chokshi1993,Poppe&Blum1997}.
The shattering and coagulation timescales are calculated following \citet{Granato21} and \citet{Aoyama17}, respectively. 
Throughout the simulation, the dust components are advected with the gas flow, behaving as passive scalars.
Readers are referred to \citet{Yohan24} for further details.

\subsubsection{Massive black hole physics}
For AGN feedback, we employ the dual mode feedback scheme including the spin-evolution of massive black holes (MBHs) \citep{2012MNRAS.420.2662D, 2014MNRAS.440.1590D, 2021A&A...651A.109D}.
We set the thermal feedback efficiency for the AGN to 0.05, which is three times lower than in \cite{2021A&A...651A.109D}. This reduces the amount of feedback energy released per unit accreted mass in quasar mode, thereby promoting black hole growth and increasing their mass at a given stellar mass, which was underestimated in previous simulations relative to observations.
The formation criterion of seed MBHs is applied based on the gas and stellar properties in each cell.
The target cell should have a gas density higher than $n_{\rm BH}=100\,\rm H\,cm^{-3}$, and a stellar mass within the cell higher than $1.88\times10^7M_{\odot}$.
This corresponds to a density of $2500\,\rm H\,cm^{-3}$ at the best spatial resolution at $z=0$.
The motivation for using the stellar mass of the cell instead of its density is to make the occurrence frequency of peak-stellar-density cells independent of spatial resolution.
Since \nc\ employs a time-dependent minimum spatial resolution, this approach helps the formation of seed black holes to be more uniformly distributed throughout the epochs.
We also require the stellar velocity dispersion to be $\sigma_* > 50\,\rm km\,s^{-1}$ and the stellar age to be $t_* > 50\,{\rm Myr}$ to avoid selecting short-lived young clumps as seed MBH formation sites.

The exclusion radius, which is the radius of the spherical region that prevents the formation of new MBH around the pre-existing MBH, is set to be $4\,\kpc$.
This constraint was originally imposed to prevent the overproduction of seeds inside a single galaxy when the formation criterion was easily met.
Compared to previous RAMSES simulations, our threshold scheme ensures the formation site of seeds to be reliably dense stellar concentrations that last long enough, making a large exclusion radius (e.g., 50 comoving kpc in  \citealp{2021A&A...651A.109D}) unnecessary. Our choice of $4\,kpc$ is enough to prevent the formation of multiple seeds near the galactic centre.

Dynamical friction of MBHs \citep{1943ApJ....97..255C,1999ApJ...513..252O} is implemented for gas from \cite{2014MNRAS.440.1590D} and stars from \cite{2019MNRAS.486..101P}.
We find that seed-mass MBHs do not settle with their own dynamical friction, even when they are near the central region of the galaxy, resulting in permanent oscillation near the galaxy centre.
This issue has also been reported in other simulations \citep{Ma+2021, Bahe+2022}.
To alleviate the problem, artificial acceleration and repositioning of the MBH have been utilised, for example, by boosting dynamical friction or the accretion rate of the MBH by a factor that depends on the gas density \citep{Weinberger+17, Wellons+23, Tremmel2018, 2021A&A...651A.109D}.

MBHs seem to show stable settling even in higher-resolution simulations that take into account dynamical interactions with both the stars and the gas components within the Bondi-Hoyle radius \citep{2019MNRAS.486..101P,2023A&A...674A.217L,2025A&A...701A.232L}.
Other simulations address stellar systems associated with the MBH (e.g., nuclear star clusters) as a critical player for the settling of the MBH \citep{2017MNRAS.469..295B, 2020MNRAS.493.3676O, 2025ApJ...981..203M, 2025ApJ...980...79Z, Shin25}.
Since our $68\,$pc spatial resolution is not enough to capture these behaviours, we apply a mild boost factor on the stellar dynamical friction of the MBH:
\begin{equation}
    f_{\rm boost} = \max[(\rho_*/\rho_{\rm crit})^{\beta}, 1]
\end{equation}
where we set $\beta=1$ and $\rho_{\rm crit}=0.16 \Msun \, \rm pc^{-3}$.
This is based on the understanding that the enhanced drag force results from unresolved high-density regions in the subgrid regime, which enables seed-mass MBHs to remain bound to concentrated stellar systems.
This binding supports their gas accretion and their early settlement at the galaxy centre by effectively deepening the gravitational potential of the system.

\subsubsection{Heating and cooling of gas}
The general radiative heating and cooling model follows the framework of \cite{2021A&A...651A.109D}.
The simulation employs a UV background from \cite{1996ApJ...461...20H} with reionisation redshift $z_{\rm reion}=10$.
Gas with density $n > 0.01\,{\rm H\,cm^{-3}}$ \citep{2012MNRAS.423..344R} has a reduced UV photo-heating rate due to self-shielding by a factor of $\exp[-n_{\rm H}/(0.01\,{\rm H\,cm^{-3}})]$.
We used tabulated cooling rates from \cite{1993ApJS...88..253S} and \cite{1972ARA&A..10..375D} that vary with the metallicity which is defined as the sum of the abundances of all gas-phase elements excluding H, He, and D.

The simulation includes an additional cooling effect from collisional interactions between electrons and dust grains at high temperature. 
We adopt the cooling rates from \cite{1981ApJ...248..138D}, which depend on the gas temperature, dust grain size, and the number densities of hydrogen and electrons, assuming that the gas is fully ionised.
The dust contribution to the gas cooling rate significantly decreases below $\sim 10^6\,\rm K$, thereby preventing dust from affecting the contraction of cold gas at lower temperatures (see \citealp{Yohan24} for details).

\begin{figure*}[!p]
    \begin{center}
        \includegraphics[width=0.9\textwidth]{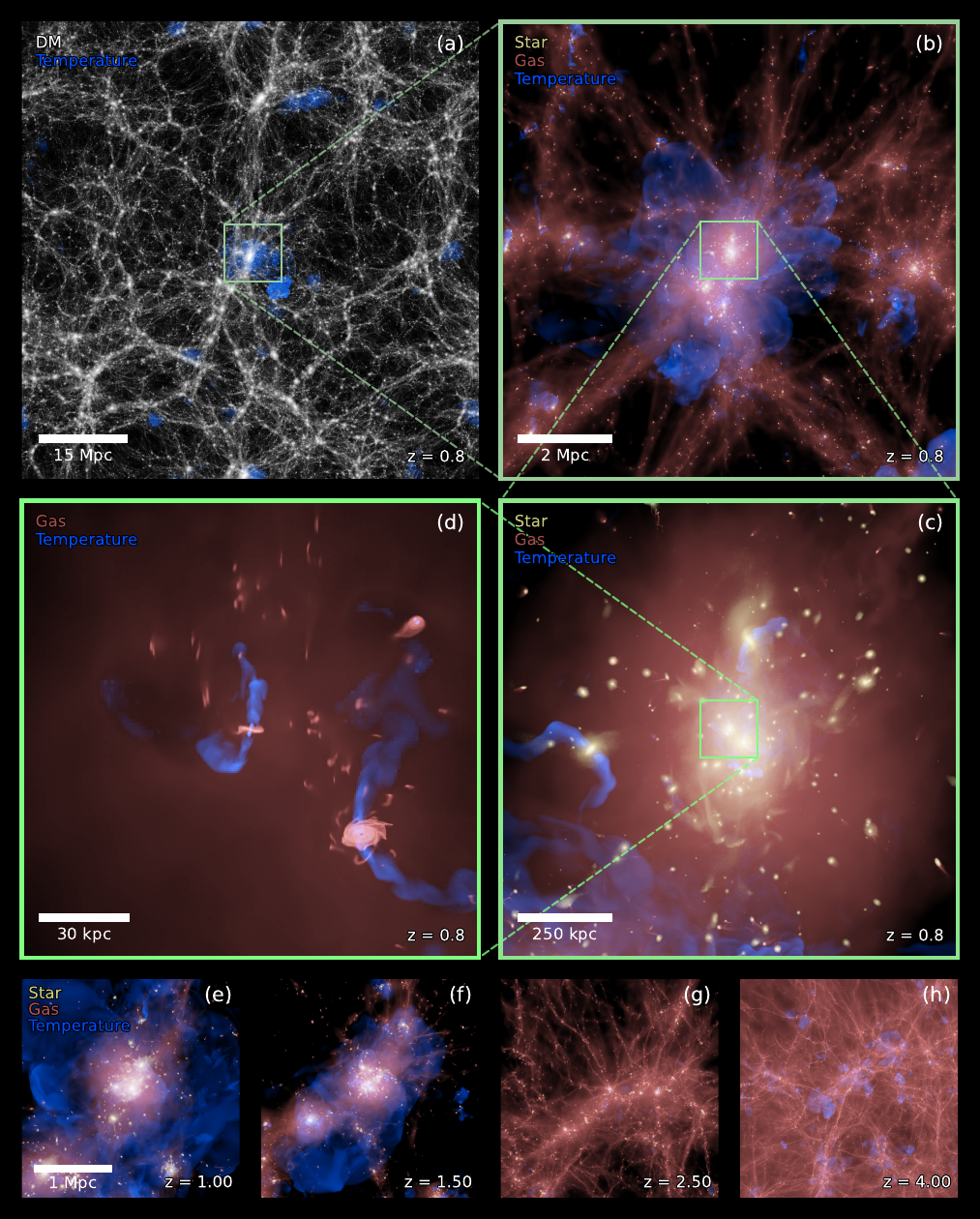}
        \caption{Overview of the \nc\ simulation. The figure shows images of (a) the large-scale volume that includes a low-resolution region, (b) the zoom-in region that includes clusters and local gas filaments, (c) the main cluster and infalling galaxies, and (d) the central brightest cluster galaxy with a relatively massive companion on the lower right side. Each panel shows a different combination of components that are specified on the upper left side with their corresponding colours. The extent of zoom-in panels (b), (c), and (d) is indicated by thin green boxes in larger panels (a), (b), and (c), respectively.
        Panels (e), (f), (g), and (h) show the region around the same target halo at different redshifts, $z=1.0$, $1.5$, $2.5$, and $4.0$.
        The scale bar indicates physical lengths in all panels.
        \label{fig:zoom_all}}
    \end{center}
\end{figure*}

\subsection{Halo and galaxy detection}
\label{section:halogaldetection}
Galaxies and halos are identified separately using the AdaptaHOP algorithm \citep{Aubert2004, Tweed2009}.
This method detects peaks and saddle points in the particle density field, which are used to estimate the initial centres and boundaries of halos.
To build a substructure hierarchy, we adopt the ``most massive sub-node method,'' which regards the most massive (leaf) substructures as the central part of the host halo while other substructures inside the host halo are classified as ``subhalos'' (see \citealp{Tweed2009} for details).
The minimum number of particles is set to 100, corresponding to $\sim10^6\, M_\odot$ and $\sim10^8\, M_\odot$ for galaxies and halos, respectively.
In total, 5,570 galaxies and 89,994 halos are identified at $z=0.8$.
However, as the zoom-in simulation contains DM particles with different resolutions, some halos are contaminated by low-resolution particles.
To consider this, we applied a strict condition that excludes halos with at least one poor-resolution DM particle, resulting in 73,745 halos in total.
The mass ranges are $10^{6.26 \text{--} 12.12}\,M_\odot$ for galaxies and $10^{8.12 \text{--} 14.15}\,M_\odot$ for halos.
The two most massive halos, which are merging at $z=0.8$, also satisfy this condition and contain no low-resolution DM particles.

Furthermore, we excluded transient and star-forming clumps from our galaxy sample.
Such clumps are often identified as substructures within galaxies, lack both longevity and clear visual distinction from the main galaxy in their stellar distributions, and are incorporated into their main host.
These clumps are defined by the following criteria: 1. they are substructures of the host group, and 2. their main progenitor branch persists for fewer than 100 snapshots (i.e., $1.5\Gyr$) or they appear as substructures in over 90\% of all snapshots.

Although AdaptaHOP computes several main halo properties such as virial radius\footnote{In AdaptaHOP, it directly computes the virial condition using member particles and defines the virial radius ($R_{\rm vir}$) where inside particles physically satisfy the condition.} or spin parameter, we further calculated key quantities related to stellar or gas components.
To measure these quantities, we apply different strategies for stellar and gas properties.
For stellar properties of a galaxy (e.g., SFR or metallicity), we only consider member particles identified by the galaxy finder.
On the other hand, gas properties (e.g., gas mass) are measured using several radius cuts, such as the stellar half-mass radius $R_{\rm 50,3D}$ (or $R_{\rm 90,3D}$), the 3-dimensional radius within which a galaxy contains 50\% (90\%) of its stellar mass, and $R_{\rm max}$, the distance from the centre to the farthest member particle.
We publicly provide all value-added catalogues on the simulation website.\footnote{https://gemsimulation.com}

\section{Results}\label{sec:results}
\subsection{Overview of the simulation}

The simulation has passed $z=0.8$ which corresponds to $6.9\,{\rm Gyr}$ after the Big Bang.
In total, 459 valid snapshots were exported with time intervals of $15\,{\rm Myr}$.
At $z=0.8$, the zoom-in region includes 19 galaxies with $M_* > 10^{11}\Msun$, 170 galaxies with $M_* > 10^{10}\Msun$, and 801 galaxies with $M_* > 10^{9}\Msun$.
The main cluster has a total mass of
$M_{\rm 200} = 1.2\times10^{14}\Msun$, and is in the beginning stage of a merger with a smaller halo of $M_{\rm 200} = 2\times10^{13}\Msun$.
This is a major merger, as the mass ratio was 1:3.2 when the secondary halo had its peak mass at $z=0.91$. 
Figure~\ref{fig:zoom_all} presents the images of the simulation volume at four different scales: (a) the entire simulation box including low-resolution regions, (b) the zoom-in region, (c) the most massive galaxy cluster, and (d) the central galaxies of the target galaxy cluster.
The colour scheme represents the various components of the simulation, as shown by the legends at the top left of each panel.
The figure also shows the same region at four additional redshifts, $z=1.0$, $1.5$, $2.5$, and $4.0$, in panels (e) through (h), each rendered at a fixed physical scale.
These panels reveal the development of large-scale structure, with galaxies as building blocks, that eventually merge into the central halo to form a cluster, as well as the hot gas bubbles and reservoirs generated by stellar and AGN feedback.

\begin{figure*}[!p]
    \begin{center}
        \includegraphics[width=1.0\textwidth]{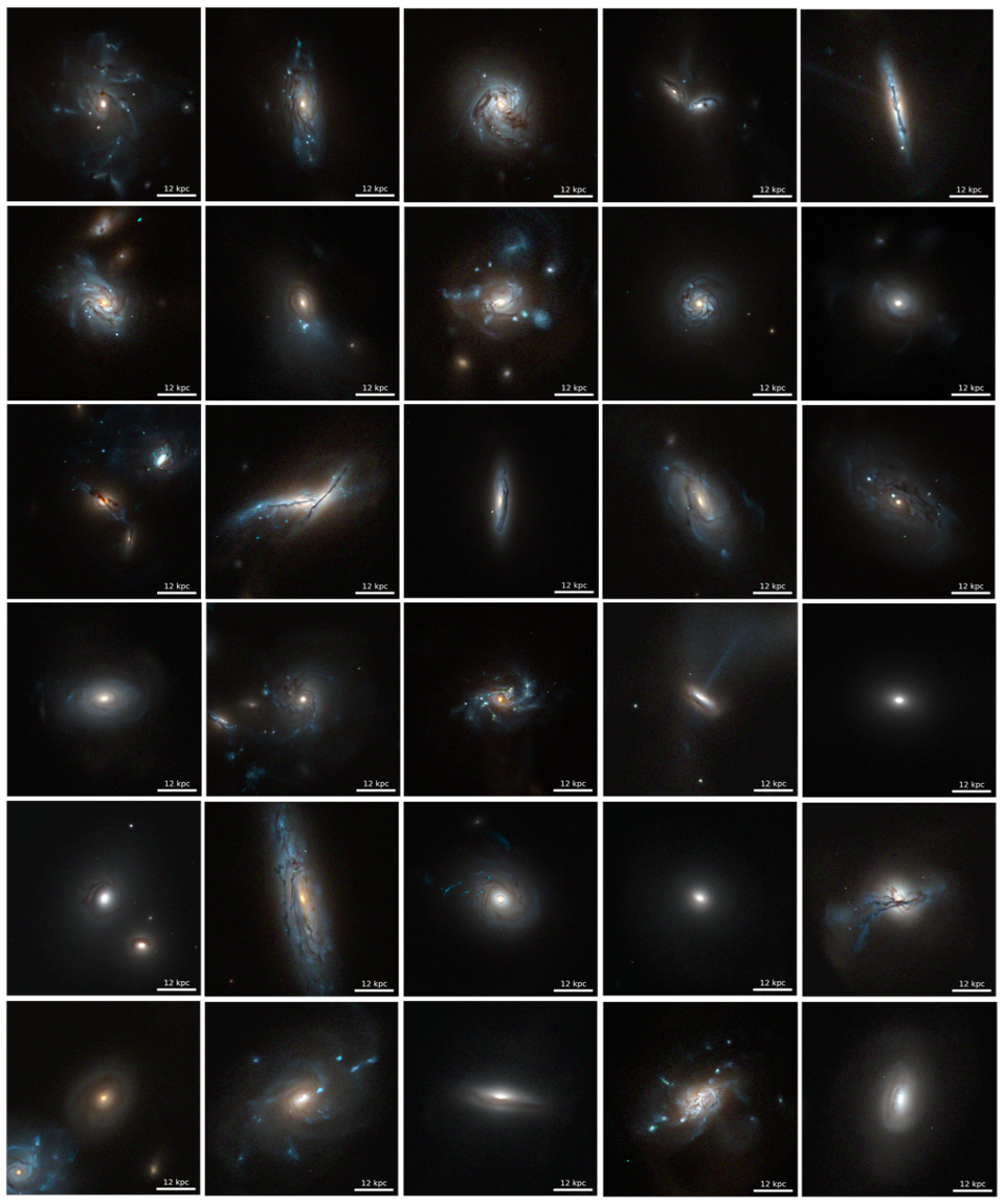}
        \caption{
        A sample of mock images of galaxies at three different redshifts ($z = 1.50$, $1.15$, and $0.80$) is shown.
        The SDSS $g$-, $r$-, and $i$-band fluxes (mapped to blue, green, and red, respectively) are used to construct the colour images.
        Based on stellar population and dust information, \nc\ successfully reproduces a wide variety of galaxy types.
        }
        \label{fig:mock}
    \end{center}
\end{figure*}

We also present the mock images of randomly selected galaxies at three different redshifts (1.5, 1.15, and 0.8) in Fig.~\ref{fig:mock}. We used the Sloan Digital Sky Survey (SDSS) $r$-/$g$-/$u$-band fluxes to create the RGB composite images.
We can clearly notice the diversity of galaxy morphologies in \nc\, especially with attenuation effects from the on-the-fly calculated dust.

\subsection{Galaxy scaling relations}
\begin{figure*}[!p]
    \begin{center}
        \includegraphics[width=1.0\textwidth]{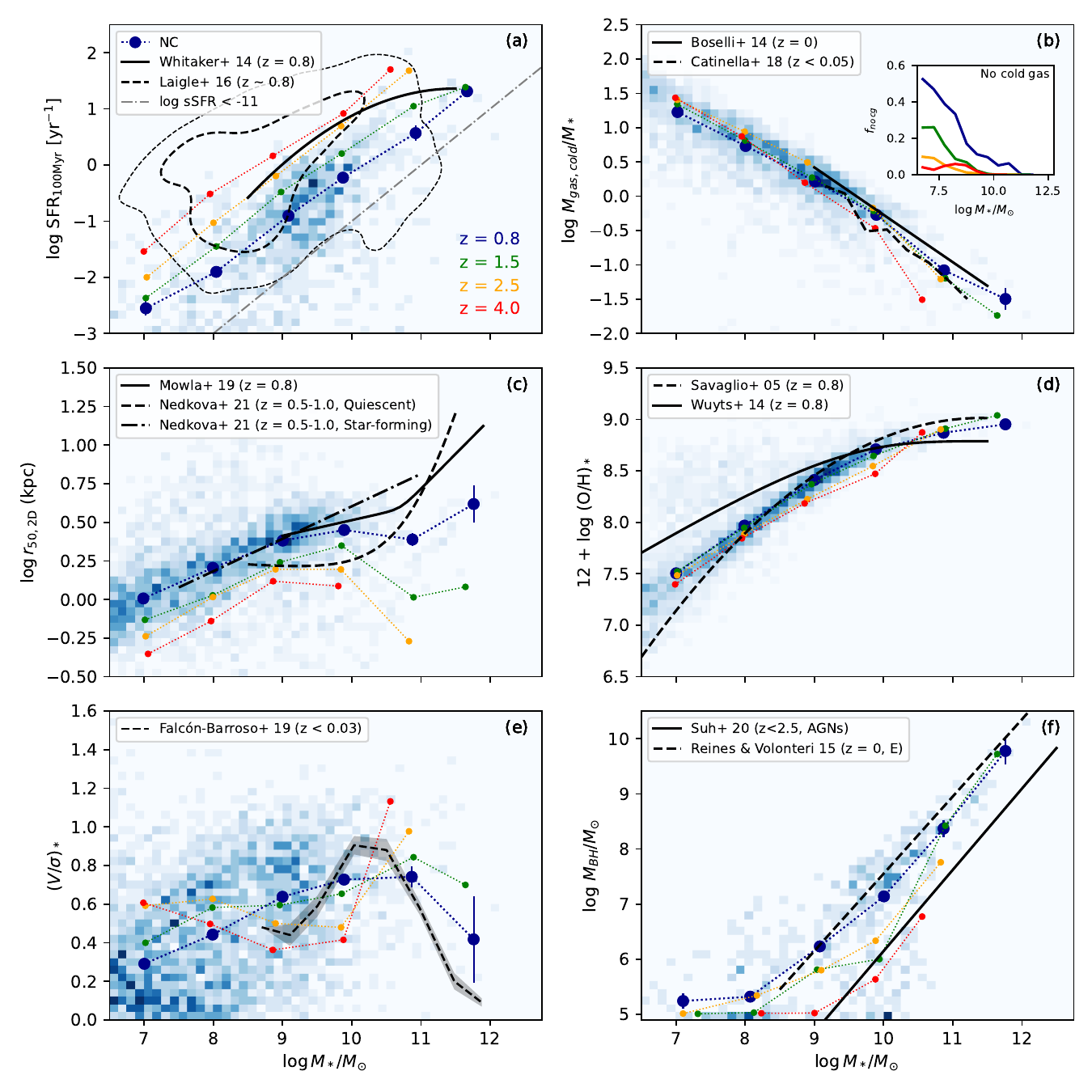}
        \caption{
        The scaling relations of \nc\ galaxies measured at $z=0.8, 1.5, 2.5$, and $4.0$.
        The $x$-axis shows the stellar mass, and the $y$-axis in each panel indicates different scaling properties for galaxies.
        The pixels on the background represent the distribution of \nc\ galaxies at $z=0.8$.
        Panel~(a) presents the star formation rate of galaxies, which shows an overall decrease with redshift.
        Panel~(b) presents the cold gas fraction of galaxies, which shows no strong evolution with redshift.
        However, the embedded figure in the panel, which presents the fraction of cold gas-deficient galaxies, shows an increase with redshift, particularly among lower-mass galaxies.
        Panel~(c) presents half mass radius, which shows an increase over time over the entire mass range.
        Significant size growth is observed in high-mass galaxies, driven by mergers followed by wet compaction events.  
        Panel~(d) presents stellar metallicity, which exhibits a tight relation consistent with the empirical trend and shows no strong evolution with redshift. 
        Panel~(e) presents rotation velocity over velocity dispersion.
        Massive galaxies initially preferentially develop a strong rotation-supported system, which can be considered as disc settling, but experience a significant decrease in their rotation speed over time, indicating the formation of a slow rotator population through mergers.
        Panel~(f) presents the mass of the most massive black holes in galaxies.
        The scaling relation lies well between the empirical relation of broad-line AGNs (solid line) and elliptical galaxies (dashed line).
        \label{fig:all_mstar}}
    \end{center}
\end{figure*}

In Fig.~\ref{fig:all_mstar}, we present some key scaling relations of galaxies from the \nc\ simulation at four different redshifts, $z=0.8, 1.5, 2.5$, and $4.0$.
Each panel shows different galaxy properties on the $y$-axis as a function of their stellar mass on the $x$-axis.
The black lines indicate empirical relations from observations, while the background pixels show the distribution of \nc\ galaxies at $z=0.8$. 
The circles indicate median values of \nc\ galaxies at each snapshot, the standard error of the mean is shown as vertical lines only for $z=0.8$ snapshot.
Panel~(a) shows the SFR versus stellar mass relation in comparison with observed main sequence from \cite{2014ApJ...795..104W}.
The distribution of galaxies at $z\sim0.8$ from \cite{2016ApJS..224...24L} is also shown with $1\sigma$ and $2\sigma$ distributions as solid and dotted contours, respectively.
SFR is measured based on the stars that are recognised as members of each galaxy.
An overall decrease in the SFR sequence of \nc\ galaxies is observed with decreasing redshift.
The sequence at $z=0.8$ shows SFR lower than the empirical relation. 
This could be attributed to the fact that the zoom-in region lies in a highly over-dense environment, where galaxies form earlier, and environmental quenching becomes effective.
As a result, the galaxy population is significantly older than that of the average field.  

Panel~(b) shows the fraction of cold gas in galaxies as a function of stellar mass.
The cold gas mass is measured by summing the gas mass of cells with temperatures below $10^{4}\,\rm{K}$ within a radius of $2\,R_{50,3D}$ from the galactic centre. 
Each median curve for the simulated galaxies has been measured for galaxies containing cold gas.
The observational relations are from \cite{2018MNRAS.476..875C} and \cite{2021A&A...648A..25Z}, where the former measured both atomic and molecular hydrogen at $z=0$, while the latter measured only atomic hydrogen gas at $z=0.8$.
The gas mass of \nc\ galaxies does not change significantly with redshift, which is consistent with the result of \cite{2021A&A...651A.109D}.
However, we find that the fraction of galaxies without any cold gas changes significantly with redshift, as shown in the embedded figure in the panel.
This indicates that gas removal occurs in certain galaxies through the violent depletion of cold gas, likely driven by environmental effects.  

Panel~(c) shows galaxy sizes, represented by the half mass radius in 2-dimension, $r_{\rm 50,2D}$.
To obtain a uniform distribution of lines of sight, we used 100 directions generated from Fibonacci lattices to measure apparent radii and used the median value of the distribution.
Empirical relations are shown from \cite{2019ApJ...872L..13M}, which includes all galaxy types (solid line), and from \cite{2021MNRAS.506..928N}, which separates galaxies into quiescent (dashed line) and star-forming (dotted line) types.
The sizes of \nc\ galaxies show a moderate increase over time for lower-mass galaxies.
The decrease in size-mass relation at high-redshift could be attributed to the gas-rich compaction event that takes place on galaxies above $\sim10^{10}\Msun$ \citep{2014MNRAS.438.1870D, 2015MNRAS.450.2327Z, 2021MNRAS.505..172L}.
These high-mass galaxies later show strong redshift evolution toward larger radii, likely driven by mergers \citep{2014ApJ...788...28V}.
At $z=0.8$, the empirical relation of star-forming galaxies closely overlaps with the lower-mass sequence of the simulated galaxies in the range of $10^{8}$–$10^{10}\,\Msun$. 
The consistency seems to degrade at higher masses, while a plateau appears in the high-mass end of $10^{10}-10^{11}\Msun$.
This can also be interpreted as the population transition from extended star-forming galaxies to compact, quiescent galaxies with increasing mass, where quenching is induced by the combination of mass and environmental effects within the cluster.
Simulated galaxies more massive than $10^{11}\Msun$ show smaller sizes compared to the empirical relation.
This discrepancy is mainly due to their overly compact central stellar concentrations, which significantly reduce their half-mass radii, even though the most massive galaxies in the simulation clearly extend well beyond $\sim10\,\rm{kpc}$ in radius.

Panel~(d) presents the stellar mass versus stellar metallicity relation of \nc\ galaxies, measured in terms of [O/H], in comparison with empirical relations from \cite{2005ApJ...635..260S} and \cite{2014ApJ...789L..40W}.
The \nc\ galaxies show no strong indication of evolution in the mass–metallicity relation across redshift.
At $z=0.8$, simulated galaxies show a tight relation that is consistent with observations, representing the continuous enrichment and recycling of gas within galaxies along with star formation.
A minor population of galaxies appears in the upper-left region of the main sequence.
Most of them turn out to be satellite galaxies that have undergone severe stripping, resulting in stellar mass loss that causes them to deviate from the main relation.

Panel~(e) shows the stellar rotational velocity $V$ divided by the velocity dispersion $\sigma$ within galaxies.
\nc\ galaxies exhibit a pronounced evolution in the relation across redshift.
At high redshifts ($z = 2.5$ and $4.0$), galaxies show stronger rotation at higher masses, consistent with a disc settling scenario that preferentially occurs in massive galaxies \citep{2019ApJ...883...25P}.
At low redshifts ($z = 0.8$ and $1.5$), the most massive galaxies evolve into less rotationally supported systems, consistent with the formation of massive slow rotators through continuous mergers \citep{1981MNRAS.197..179G,1988ApJ...331..699B,CY17}.
This behaviour is consistent with \cite{Dubois16}, who found that the $(V/\sigma)_*$ trend peaks around a stellar mass of $10^{11}\,\Msun$.
The median relation derived from the observed sample of \cite{2019A&A...632A..59F} at $z\sim0$ is shown as a black dashed line, which follows a similar trend to the extrapolated redshift evolution of the simulated galaxies toward lower redshift.

Panel~(f) shows the mass of the most massive black holes within each galaxy, compared with the empirical relations of \cite{2020ApJ...889...32S} and \cite{2015ApJ...813...82R}.
The former is derived from observations of broad-line AGNs at $z<2.5$, shown as the solid line, while the latter is derived from classical bulges and ellipticals, shown as the dashed line.
Regardless of the redshift, \nc\ galaxies lie between the elliptical and AGN relations and exhibit a large scatter around the empirical trends, reflecting the morphological diversity of the sample as seen in Fig.~\ref{fig:mock}.
\nc\ galaxies show a moderate increase in MBH mass at fixed stellar mass with decreasing redshift, suggesting a possible redshift evolution in the black hole mass–host galaxy mass relation.
However, this trend may also be influenced by biases arising from the general morphological evolution of galaxies, as early-type galaxies become increasingly common at lower redshifts.

\begin{figure}[!t]
    \resizebox{\hsize}{!}{
    \includegraphics{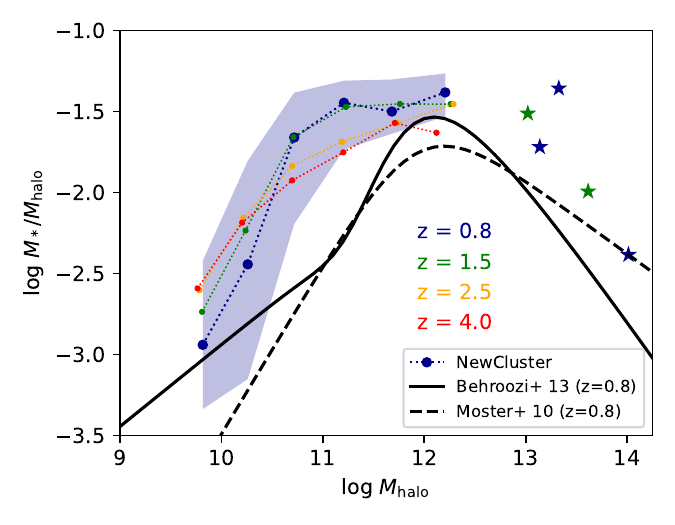}}
    \caption{
    Stellar-to-halo mass ratio as a function of halo mass for all central galaxies.
    Circles show median values in each bin, colour-coded with different redshifts.
    Shade indicates $1\sigma$ scatter of the distribution at $z=0.8$.
    Stars indicate individual galaxies.
    Empirical relations at $z=0.8$ are shown as black lines.
    \label{fig:SHMR}}
\end{figure}

Figure~\ref{fig:SHMR} shows the stellar-to-halo mass ratio as a function of halo mass for \nc\ galaxies.
The halo mass is measured at the peak mass of the main progenitor.
Circles indicate the median of the distribution in each bin, colour-coded with different redshifts.
The $1\sigma$ scatter at $z=0.8$ is shown as shades.
Stars indicate individual galaxies with halo mass above $10^{13}\Msun$.
Empirical relations are shown for \cite{2013ApJ...770...57B} as the solid line and \cite{2010ApJ...710..903M} as the dashed line.
Although \nc\ galaxies lie in a high-density environment that may exhibit a stellar-to-halo mass ratio different from the universal relation, the excess of stellar mass in the intermediate-mass halos ($M_{\rm halo} \sim 10^{11}\Msun$) may indicate the over-production of stars for dwarf galaxies.
This suggests that the current stellar feedback processes are not effective at suppressing star formation in low-mass galaxies at high redshifts.
However, their effects at $z < 1$ appear to lead to an overly quenched population of galaxies, as shown in Panel~(a) of Fig.~\ref{fig:all_mstar}.

\begin{figure}[!t]
    \resizebox{\hsize}{!}{
    \includegraphics{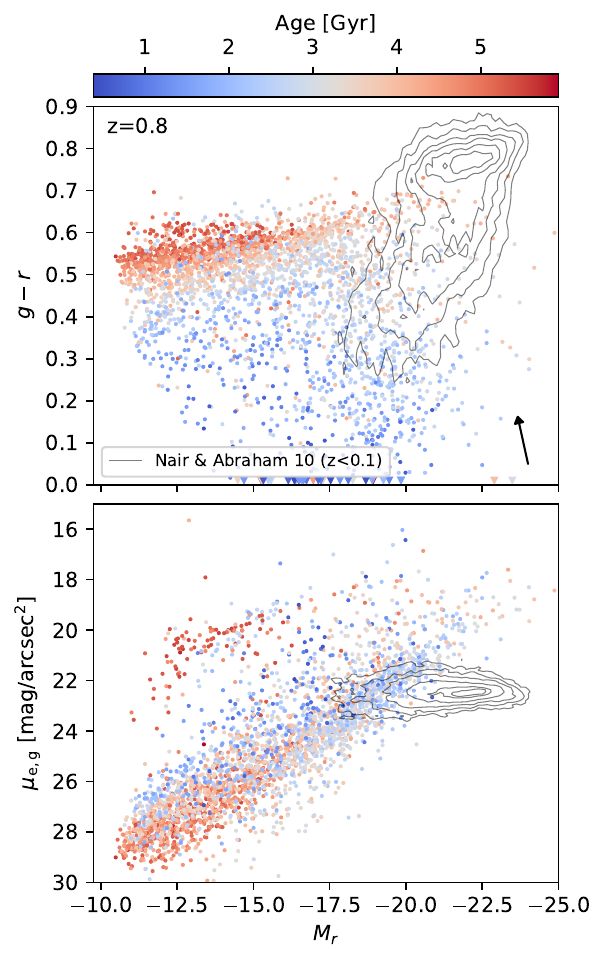}}
    \caption{
    Photometric properties of \nc\ galaxies at $z=0.8$.
    The $x$-axis shows the $r$-band absolute magnitude.
    The $y$-axis of the top (bottom) panel represents $g-r$ colour ($g$-band effective surface brightness).
    We also present the black arrow to show the effect of dust attenuation assuming $\rm E(B-V)=0.1$ with Calzetti attenuation curve \citep{Calzetti2000}.
    The colour of data points indicates the mass-weighted age of galaxies.
    The grey contours in both panels represent local observations from SDSS \citep{Nair2010}.
    \label{fig:CMD}}
\end{figure}

We present the photometric properties of \nc\ galaxies at $z=0.8$ in Figure~\ref{fig:CMD}.
The sample galaxies are the same as those in Fig.~\ref{fig:all_mstar}, but excluding misidentified clump-like substructures.
We measure the $g$- and $r$-band absolute magnitudes using stellar age and metallicity, based on the simple stellar population models from \cite{CB07}.\footnote{We simply measure the photometric magnitudes from the averaged age and metallicity of stars following the table, not considering dust absorption.}
The top panel shows the $g-r$ colour versus $r$-band absolute magnitude, while the bottom panel presents the $g$-band surface brightness within the effective radius.
Different colours indicate the $r$-band-weighted age of galaxies.

The observed distributions of local galaxies are also displayed \citep{Nair2010}.
Although the overall and median (green error bar) locations are not identical to observations, the trends are consistent with naive expectations.
For instance, our colour-magnitude distribution shows the separation between old and young populations, while the red sequence is not clearly separated at this redshift.

In the bottom panel, central surface brightness also linearly correlates well with absolute magnitude, as reported in observations of the Local Group \citep{McConnachie2012}.
We note that a separate cluster of objects is observed in the upper left of the main sequence.
Their old age, low luminosity, and high surface brightness indicate that these objects are compact and quenched.
After tracing their evolution through the merger tree, we find that they are not transient structures, as they show long, continuous histories without breaks. 
Since their progenitors are usually found as substructures within larger galaxies in terms of spatial distribution, they are likely ultra-compact dwarf galaxies (UCD) or orphan stellar systems that were ejected from larger galaxies \citep[e.g.,][]{Jang2024}.
We leave the detailed investigation of these galaxies to future work.

\begin{figure}[!t]
    \resizebox{\hsize}{!}{
    \includegraphics{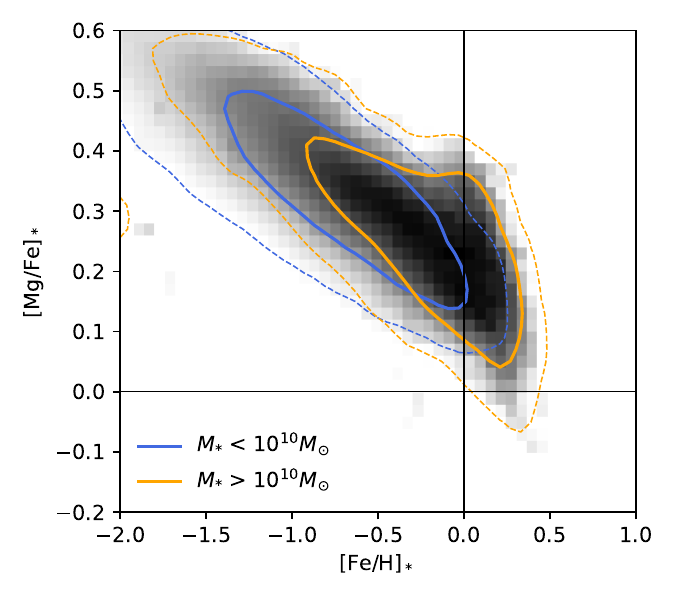}}
    \caption{
    Distribution of stars on the [Fe/H] - [Mg/Fe] chemical composition plane for galaxies in \nc\ at $z = 0.8$. The grey 2-dimensional histogram in the background indicates the distribution of all stars. Contours indicate stars which are members of galaxies with different stellar masses drawn by contours that contain $68.2\%$ (solid lines) and $95.5\%$ (dashed lines) of their distribution.
    \label{fig:alpha_chem}}
\end{figure}

Figure~\ref{fig:alpha_chem} shows the distribution of the stellar alpha enhancement fraction ([Mg/Fe]) as a function of metallicity ([Fe/H]).
The grey background indicates the entire population of stars in the simulation, while the two contours show the members of galaxies in two different stellar mass ranges.
As reported by previous studies \citep[e.g.,][]{2014A&A...562A..71B}, the alpha‐element enhancement fraction decreases with increasing metallicity, reflecting the differential enrichment from SNII and SNIa with their respective delay‐time distributions.
Comparing different stellar mass ranges, the stellar populations in more massive galaxies show overall higher metallicities and, at a given metallicity, higher alpha-element enhancement fractions.
This indicates that the more massive galaxies in our simulation had more recycling of gas but with a shorter star formation history.

\subsection{Low-surface brightness structures}
\begin{figure*}[!t]
    \sidecaption
    \includegraphics[width=12cm]{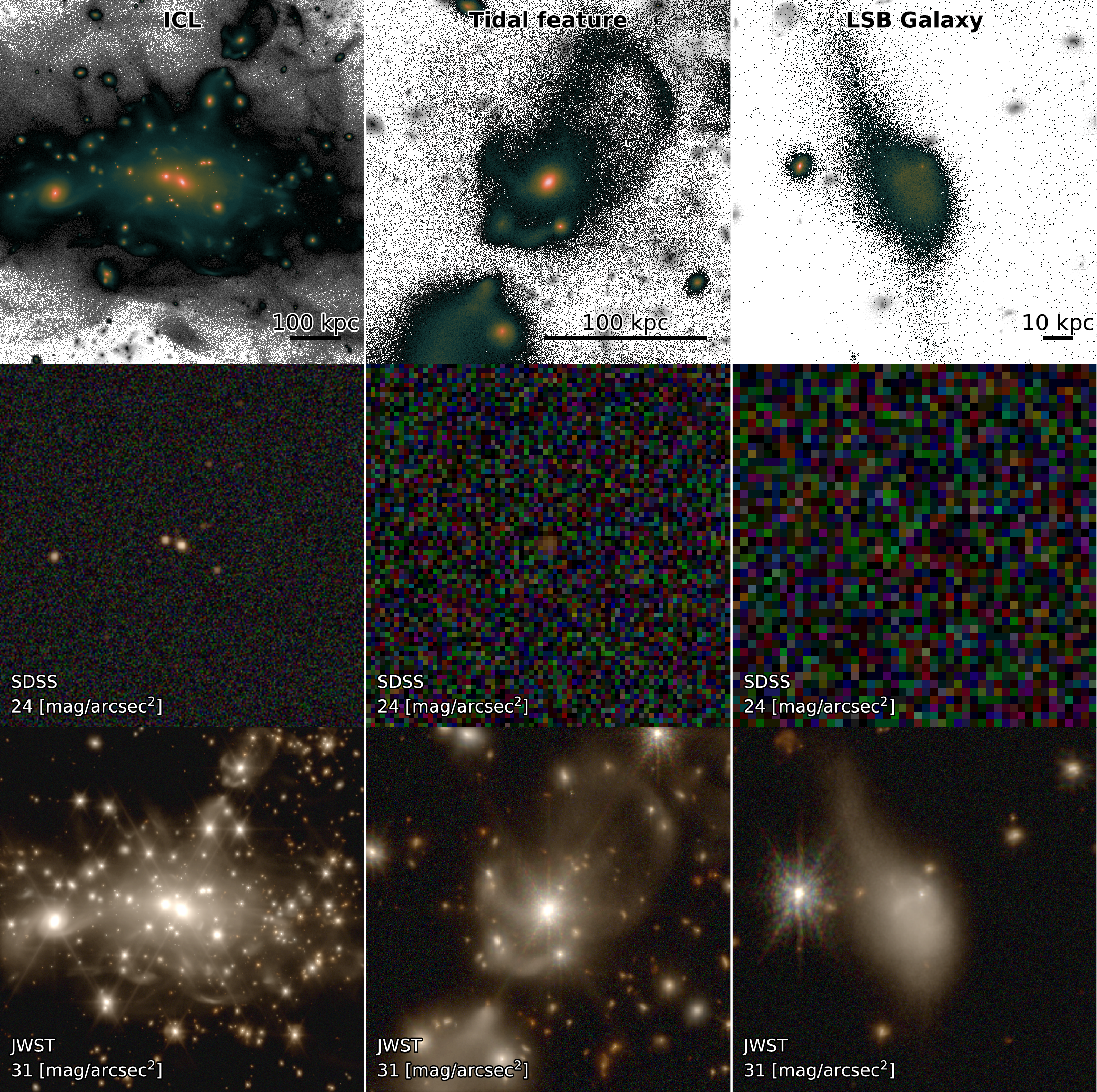}
    \caption{
    Stellar density maps (top), mock SDSS, and JWST composite images (middle and bottom) of various objects in \nc.
    From left to right, images display the brightest cluster galaxy, tidal feature, and low-surface brightness galaxy samples.
    The sources from earlier snapshots are included to express the high-z background objects, and we also consider the k-correction to generate mock observations.
    Mock SDSS and JWST images are a composite of $r$, $i$, and $z$ filters to cover the JWST range.
    We arbitrarily add a sky noise corresponding to 24 and 31 $\rm{mag\,arcsec^{-1}}$ for SDSS and JWST, respectively.
    The point spread functions are convolved with the pixelized source images.
    \label{fig:LSB}}
\end{figure*}

Low-surface brightness (LSB) structures, such as tidal streams, UDGs, and ICL, are emerging as key tracers of dynamical evolution through the assembly history \citep{vanDokkum2015, Martin2019, Contini2021, Contini2023}.
They are observed on a wide range of scales, from dwarf galaxies to massive galaxy clusters, and are generally defined as regions fainter than a surface brightness threshold of $22-24\,\rm{mag\,arcsec^{-2}}$ \citep{Bothun1997, Bakos2012, Tanoglidis2021}.
Thanks to the fine stellar mass resolution of our simulation, we are able to resolve these faint structures at $z \sim 0.8$.
LSB features of galaxies are receiving increasing attention due to their potential to shed light on the past assembly history of galaxies.
UDGs are also interesting targets for studying extreme sites of star formation and their distinct characteristics compared to typical dwarf galaxies \citep{vanDokkum2015, Pandya2018, KadoFong2022}.
Moreover, ICL has the potential to serve as an observable tracer of a cluster's dynamical evolution and even the distribution of DM \citep{Montes2019, Contini2021, Contini2023}.  

Despite being a broadly appealing target for study, LSB structures are inherently difficult to investigate due to their low detectability \citep{Contini2021, Yi2022, Joo2023, Kelvin2023, KimJ23}.
This is especially true at high redshifts, where galaxies may exhibit more dramatic tidal features because they are dynamically younger \citep{Hopkins2010, Deger2018, Ferreira2020}, but most studies are based on observations at $z < 1$ due to detection limits reduced by cosmological dimming.
The emergence of cutting-edge instruments (e.g., JWST, LSST) presents an opportunity to extend our studies to higher redshifts and is expected to bring numerous LSB detections.
It is necessary to prepare corresponding high-resolution simulations to resolve faint structures in a cosmological context in order to capture a wide range of their formation scenarios.

The minimum stellar particle mass of \nc\ is $\sim2\times10^4\,M_\odot$, which corresponds to a theoretical $r$-band surface brightness limit of $33.5\,\rm{mag\,arcsec^{-2}}$ at $z=0.8$.
This allows us to resolve faint structures on various scales.
In the top panels of Fig.~\ref{fig:LSB}, we display stellar density maps of the ICL, a tidally distorted galaxy, and an LSB galaxy from left to right.
Many faint structures are visible, not only in the ICL, but also in the trails of satellite galaxies.
As more dynamical interactions are expected among galaxies in cluster environments \citep{Dressler1980, Jaffe2011, Yi2013}, our simulation is well suited for studying the evolution of interacting galaxies down to the dwarf galaxy regime.

To enable direct comparison between our model and high-redshift observations, we also generate mock SDSS- and JWST-like images for these targets, shown in the middle and bottom panels of Fig.~\ref{fig:LSB}.\footnote{Since the wavelength coverages of SDSS and JWST are different, we arbitrarily select filters of $r$, $i$, and $z$ bands of SDSS, which correspond to nearly the F070W and F090W filters of JWST. Point spread functions (PSFs) at these JWST wavelengths are generated using the STPSF package (formerly WebbPSF; \citealp{Perrin2012}).}
We note that the mock SDSS images do not include the atmospheric seeing effect and therefore represent an idealised case compared to actual SDSS observations.
It is challenging to capture the LSB features at $z\sim0.8$ with SDSS-like instruments (middle row).
In particular, LSB galaxy is not visible in the SDSS mock image due to limited resolution and sensitivity, even though we select a relatively massive ($M_*=3.5\times10^9\Msun$) galaxy.
In contrast, applying the resolution and PSF of JWST (bottom row) enables the detection of faint structures seen in the stellar density map (top row).

A detailed investigation of the origins of the LSB features presented here is beyond the scope of this work.
However, our results demonstrate that the simulation is well suited for direct comparison with high-redshift observational studies of LSB structures, including analyses of their detailed properties and evolutionary histories.
The dense output cadence of $\sim15\,{\rm Myr}$ allows us to trace the detailed motions of individual stellar particles stripped from galaxies.
Moreover, by tracking both chemical and dust evolution, the model enables direct comparison with spectroscopic observations.
Our simulation has not yet reached low redshifts, where LSB observational data are more abundant.
Nevertheless, the current snapshot remains valuable not only for direct comparison with observations at $z \sim 0.8$ but also for investigating the dynamical evolution of stars across a wide range of spatial scales.

\subsection{Gas duty cycles and AGN feedback}
\begin{figure}[!t]
    \centering
    \resizebox{0.8\hsize}{!}{
    \includegraphics{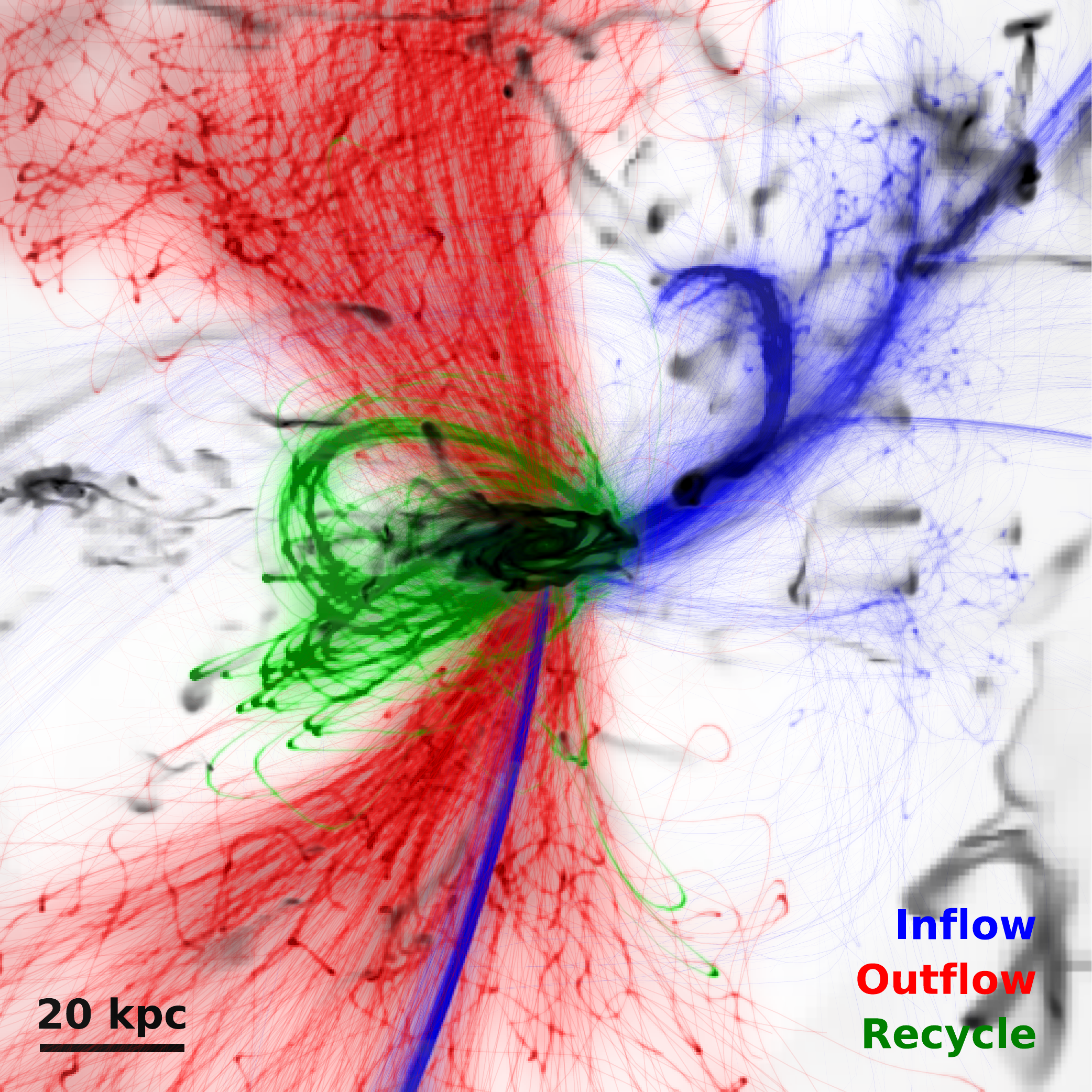}}
    \caption{
    Smoothed trajectories of gas tracer particles that are located in the central region of the galaxy at $z=1.6$.
    The greyscale background represents the gas temperature (i.e., cold gas in a darker colour).
    Blue lines are the trajectory of the inflow gas 500\, Myr before the central AGN activity.
    Red lines indicate outflow gas expelled by the violent jet.
    Green lines are similarly blown out but return back to the galactic disc.
    \label{fig:tracer}}
\end{figure}

Gas motion is intrinsically complex, governed not only by secular hydrodynamics but also by energetic astrophysical events such as SN explosions and AGN outflows.
\nc\ is performed via AMR technique, an Eulerian approach that excels at capturing shocks and discontinuities during the feedback processes \citep{2002A&A...385..337T, O'Shea2005, Hubber2013}.
However, despite this strength, AMR does not track the motion of individual gas cells, limiting its ability to follow Lagrangian gas trajectories, which is important for understanding the impact of feedback processes \citep{Bellovary2013, Choi2024}.
In galaxy evolution studies, this makes it difficult to distinguish between inflowing and outflowing gas, which is crucial for understanding the cold gas supply and the impact of feedback on star formation.

To overcome this limitation, we apply Monte-Carlo tracer particles that follow the hydrodynamics of gas cells \citep{Genel2013, Cadiou2019}.
\citet{Cadiou2019} demonstrated that Monte-Carlo tracers exhibit better performance than velocity-advected models in tracing the gas motion and spatial distribution in cosmological simulations.
The simulation includes 267,997,944 tracer particles in total.
These tracer particles probabilistically move between cells and stars, and massive black hole particles based on their gas mass exchange events, such as flux between cells, star formation, mass ejection from feedback, and accretion.
Tracer particles provide valuable opportunities by allowing the tracking of actual gas motions via particle IDs within the simulation framework.
We found that the spatial distribution of gas tracers closely resembles the density distribution of gas cells within the simulation.
However, we note that the discrete number of tracer particles cannot perfectly reproduce the true transport of gas mass, a limitation shared with the full Lagrangian approach.

In Fig.~\ref{fig:tracer}, we show the trajectories of gas tracer particles within a galaxy as an example of their use in tracking gas dynamics.
The central galaxy ($M_*=2.2\times10^{12}\,\Msun$ in $M_{\rm host}=4.63\times10^{13}\,\Msun$) experiences a violent AGN outburst at $z=1.6$. 
We first selected sample gas tracers that were initially located within $R_{\rm 90,3D}$ of the central galaxy and further classified them into three categories based on their trajectories before and after $500\,\mathrm{Myr}$:
\begin{enumerate}
    \item Inflow: The time when the tracer is farthest from the central MBH occurs before reaching its pericentre, and its maximum distance exceeds $10\,R_{\rm 90,3D}$.
    \item Outflow: The tracer reaches a maximum distance greater than $6\,R_{\rm 90,3D}$ and continues to move away from the centre for at least $400\,{\rm Myr}$.
    \item Recycle: The tracer travels outward beyond $2.5\,R_{\rm 90,3D}$ but later returns within $1.5\,R_{\rm 90,3D}$.
\end{enumerate}
We show the smoothed trajectories of individual gas tracer particles, classified into each category and colour-coded accordingly.
The background displays the gas temperature.  
Since the target galaxy is moving from left to right, most of the inflowing gas (blue) originates from the right side of the figure.  
However, it is clearly visible that some gas inflow also originates from the bottom, exhibiting a narrowly-channeled feature that indicates gas brought in by an infalling satellite galaxy.
We note that the theoretical study reported that the most stream influx comes from one or few dominant streams \citep{Danovich2012}.
Outflowing tracers (red) rapidly escape from the galactic centre along the bipolar AGN jet axis, which is aligned with the galaxy’s rotation axis.
They also show highly turbulent motion after reaching the low-density circumgalactic region.
Recycled tracers (green) do not display a strongly preferential direction, in contrast to inflow or outflow tracers.
The presence of recycled gas demonstrates that a significant amount of gas can escape from the galactic disc and later return, consistent with the galactic fountain scenario \citep{Bregman1980, Melioli2008}.

Tracer particles are directly applicable to investigate how AGN feedback ejects galactic gas and regulates central star formation.
Moreover, as our simulation is set in a cluster environment, it is also suited to studying how gas in satellite galaxies is stripped in ICM by ram pressure or how their gas supply is halted.
Tracer particles also provide a means to study the MBH duty cycle \citep{Martini2001, Choi2024}.
Since we track the state transition of tracer particles to gas, star, and MBH tracers, we can distinguish the gas consumed by accretion from that expelled by feedback.

\subsection{Ram-pressure stripping of gas}

\begin{figure*}[!t]
    \sidecaption
    \includegraphics[width=12cm]{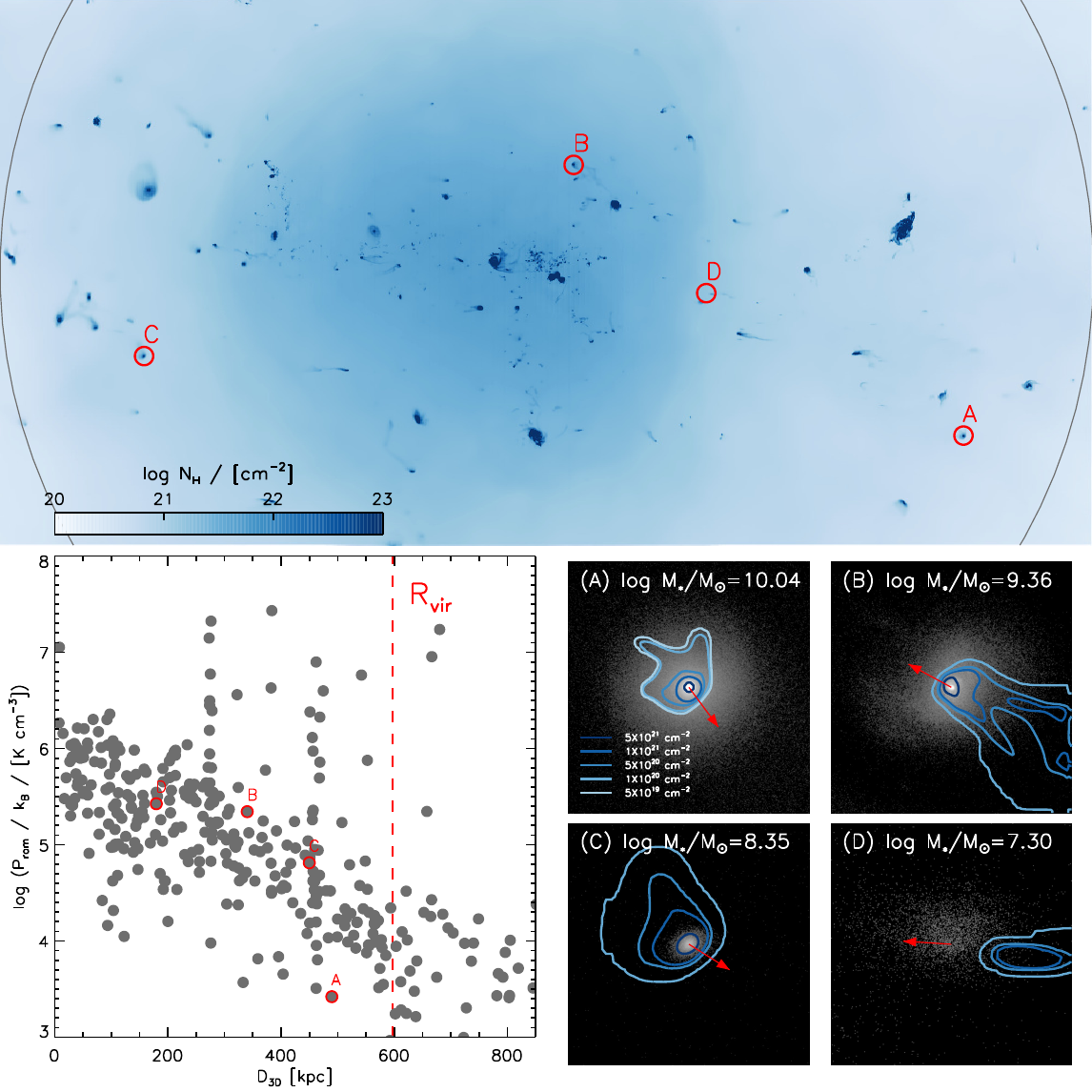}
    \caption{
    Upper panel: Gas column density map of the main cluster in the \nc\ simulation at $z = 1$. 
    The virial radius of the cluster is drawn with a grey circle.
    Numerous galaxies show gas tails within the early-stage cluster.
    Four galaxies chosen for representative illustration are marked with red circles.
    Bottom panel (left):
    Radial profile of the ram pressure exerted on galaxies (grey circles).
    The red dashed vertical line shows the virial radius.
    An evident decreasing trend of the magnitude of ram pressure can be seen.
    The ram pressures of the four representative galaxies are shown with the red circles.
    Bottom panel (right):
    Stellar density and gas column density distributions of the four galaxy samples.
    In each panel, the background density map is the surface brightness of the galaxy in $r$-band, and the contours show the column density distributions of cold gas.
    The red arrow indicates the direction of motion of the galaxy.
    }
    \label{fig:rps}
    
\end{figure*}

Cluster environments host the most significant gas removal processes observed in galaxies.
The interaction between the hot ICM and the ISM of galaxies moving at velocities of $\sim 1000\,{\rm km\,s^{-1}}$ exerts strong pressure, estimated as $P_{\rm ram} / k_{\rm B} \sim 10^{5-6}\,{\rm K}\,{\rm cm}^{-3}$ \citep[][]{Jung18, Yun19, Boselli21}, following the formulation of \cite{GG72}.
This external pressure often exceeds the internal gravitational restoring force of galaxies, leading to the stripping of their gas components.
Observations of gas trails including jellyfish phenomenon~\citep[e.g.,][]{Kenney04, Chung09, Jaffe18} provide evidence for this mechanism.
The resulting gas loss effectively quenches star formation (SF), often on short timescales although a temporal enhancement of SF activity occurs due to compression of ISM \citep[e.g.,][]{Crowl06, Vulcani18,Roberts20}.
Consequently, strong ram pressure, in combination with other environmental mechanisms, is now recognised as a driver of the high fraction of passive galaxies observed in clusters \citep[e.g.,][]{Contini20, Rhee20, Oman21}.

An important next step is to understand how complex the interactions between the ICM and ISM truly are, and how gas is stripped from galaxies in a manner that reproduces the observed morphologies.
One effective approach is the use of idealised wind-tunnel simulations, where a uniform ICM wind is applied to a model galaxy to mimic ram pressure stripping \citep[e.g.,][]{Bekki09, Tonnesen12, Lee20,Choi22}.
These simulations offer very high spatial and temporal resolution, enabling detailed studies of gas dynamics under controlled conditions.
However, a key limitation is that the ram pressure strength is often assumed to be constant, which does not reflect the time-varying conditions experienced by galaxies in live halos \citep[c.f,][]{Tonnesen19}.
Recent studies suggest that such variations in ram pressure may play a role in shaping both the gas content and morphology of galaxies \citep[e.g.,][]{Samuel23,Zhu24}.
Therefore, it is essential to incorporate realistic ram pressure profiles when studying its effects on galaxies.

Cosmological hydrodynamic simulations provide a complementary framework for studying the impact of ram pressure on galaxy evolution.
For example, \cite{Jung18} used the YZiCS simulation \citep{CY17} to investigate gas removal in clusters, finding that strong ram pressure near cluster centres is a dominant driver of gas deficiency in satellite galaxies.
Several other studies employing cosmological simulations \citep[e.g.,][]{Lotz19, Yun19, Rohr23} have similarly concluded that ram pressure plays a central role in driving environmental quenching in clusters.
However, these cosmological simulations often lack the spatial resolution necessary to capture the complexity of ICM-ISM interactions, particularly the multiphase or clumpy nature of ISM, which may be more resilient to stripping.
As a result, there is significant potential for overestimating or overlooking ram pressure effects in cosmological simulations, especially when small-scale physical processes fall below the resolution limit.

The \nc\ simulation bridges the gap between idealised and cosmological approaches to studying ram pressure stripping.
Its spatial resolution is sufficient to capture high-density gas components---possibly corresponding to the molecular phase of ISM---that are expected to be less sensitive to ram pressure \citep[e.g.,][]{Tonnesen09}.
In addition, the mass resolution enables the modeling of ram pressure effects on dwarf galaxies with $M_{*}>10^{7-9}\,M_{\odot}$, which is a mass range that remains largely inaccessible to most large-scale simulations.
Although \nc\ does not extend to the local Universe, it captures the high-redshift regime ($z>0.8$), a period when ram pressure was already a significant environmental process \citep[e.g.,][]{Simons20}.
As such, the simulation offers an opportunity to investigate the early formation of gas-poor galaxies in dense environments and to shed light on the onset of environmental quenching in nascent clusters.

The upper panel of Fig.~\ref{fig:rps} illustrates the gas distributions within the main cluster in \nc\ at $z = 1$, where the gas column density is colour-coded and the virial radius of the main cluster is shown as the grey circle.
Although the cluster has not yet reached the canonical mass range for local clusters ($M_{\rm vir} > 10^{14}\,M_{\odot}$), numerous galaxies display extended gas tails, indicating that ram pressure effects are already active at $z = 1$.
To evaluate whether ram pressure is sufficiently strong to influence galaxy evolution, we compute its value for individual galaxies surrounding clusters.
The volume-weighted mean ram pressure is measured using ICM gas cells, following the methodology of the ICM gas selection in \cite{Rhee24}:
\begin{equation}
     P_{\rm ram} = 
     \frac{\sum\limits_{i} \rho_{i} |\vec{v}_{i}|^{2} V_{i} }{\sum\limits_{i} V_{i}},
\end{equation}
where $\rho_{i}$, $\vec{v}_{i}$, and $V_{i}$ are the density, velocity with respect to the galaxy, and the volume of $i$-th gas ICM cell, respectively.
The bottom left panel of Fig.~\ref{fig:rps} presents the radial profile of ram pressure across the galaxy sample.
A clear decline in ram pressure with increasing distance is seen, primarily reflecting the decreasing ICM density at larger radii.
The scatter in this trend is likely driven by variations in galaxy velocities.
In addition, galaxy–galaxy interactions can also induce strong ram pressure.
For example, galaxies experiencing strong ram pressure at around $D_{\rm 3D} = 250$ or $450\,{\rm kpc}$ are satellites undergoing significant ram pressure from the gas disc of a massive galaxy.
Notably, galaxies located near the cluster core experience ram pressure values on the order of $P_{\rm ram}/k_{\rm B} \sim 10^{6}\,{\rm K}\,{\rm cm}^{-3}$, consistent with estimates for local clusters \citep[][]{Jung18, Yun19, Boselli21}.

The bottom-right panels of Fig.~\ref{fig:rps} display four typical galaxies with different stellar masses undergoing ram pressure stripping, each of which is also marked in the cluster gas density map with the red circles in the upper panel.
In each panel, the background shows the stellar surface density, while overplotted contours trace the column density of cold gas, with different contour levels with different colours.
The red arrow indicates the direction of motion of the galaxies.
All four galaxies exhibit asymmetric gas morphologies, with cold gas displaced in the direction opposite to their motion---a characteristic signature of ram pressure stripping.
Notably, Galaxy D, which has the lowest stellar mass and experiences the strongest ram pressure among the four, shows complete removal of its cold gas, illustrating the dramatic impact of ram pressure stripping in low-mass galaxies.

Therefore, the \nc\ simulation is useful to study the impact of ram pressure in early clusters on satellite galaxies across a wide range of stellar mass.
Comparing the measured strength of ram pressure with the restoring force inside galaxies provides quantitative investigation of the origin of gas stripping.
Determining whether galaxies undergoing ram pressure retain their gas and continue SF, or are fully quenched, remains an important topic for future studies.
The use of tracer particles also enables the study of the distribution of stripped gas inside cluster halos, but we defer such a detailed investigation to future work.

\subsection{Dust properties}

\begin{figure*}[!t]
    \centering
    \includegraphics[width=17cm]{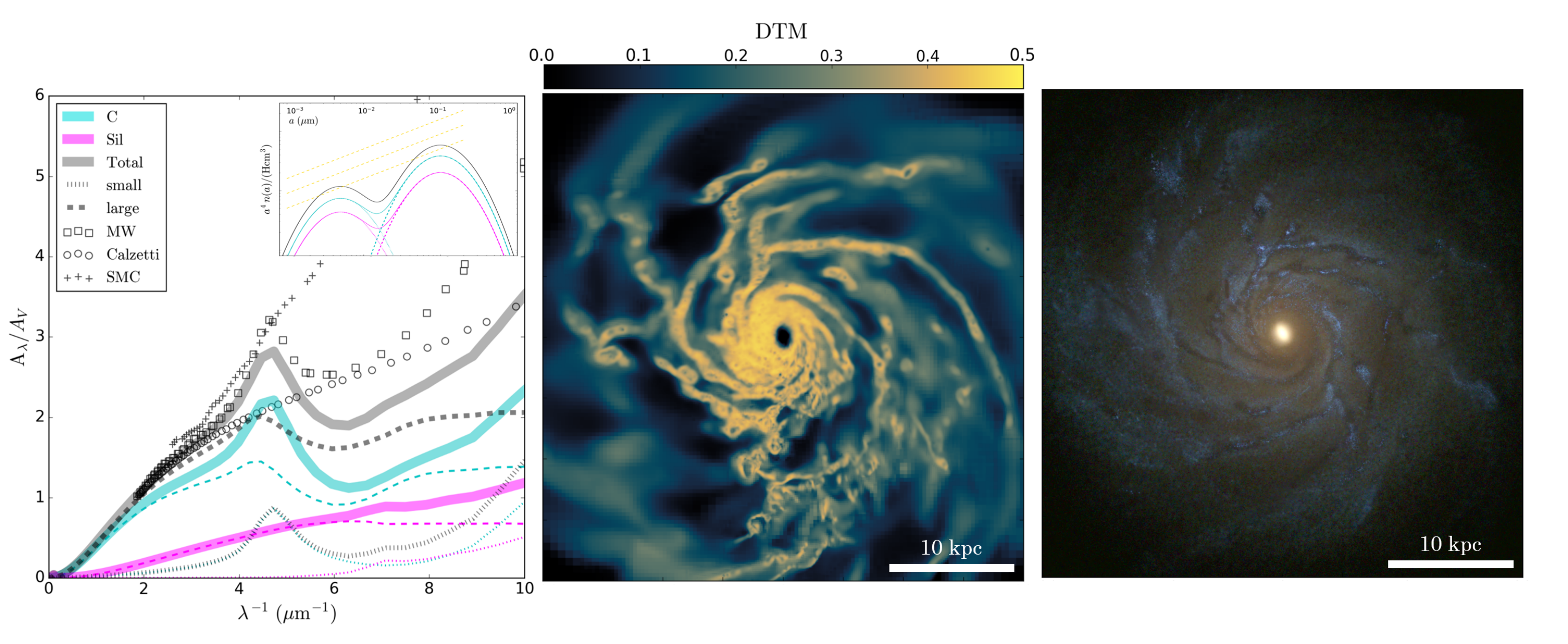}
    \caption{An example galaxy extinction curve and corresponding mock observation. The extinction curve is shown in the first column of the figure. The total extinction curve (grey) is presented along with the contributions from graphite (cyan) and silicate (magenta) grains, each shown with two different size distributions (dotted lines for small grains and dashed lines for large grains). The MW \citep{Fitzpatrick&Massa2007} extinction, SMC \citep{Pei1992} extinction, and Calzetti attenuation curves are also included for comparison. The grain size distributions for different grain types are shown in the sub-panel, with the MRN \citep{Mathis1977} slope indicated by a yellow dashed line. The spatial distribution of the DTM is shown in the second column. The corresponding mock image, created using SDSS g, r, and i bands for RGB, is shown in the last column.
    \label{fig:dtm_grain}}
\end{figure*}

The JWST presents the unprecedented data of the early Universe.
Understanding the data requires a proper knowledge of dust properties and of its evolution because dust changes the intrinsic spectral energy distribution (SED) of galaxies significantly.

Numerical simulations have recently begun to follow dust evolution in a cosmological context on the fly, including dust formation, destruction, and the evolution of chemical composition and grain size \citep{Dave2019, Jones2024, Choban2025}.
This approach goes well beyond the simpler post-processing methods that map dust onto the metal mass content of the gas.
Since \nc\ tracks the evolution of carbonaceous and silicate grains separately, it enables a detailed examination of the distinct behaviours of different dust species as they respond to dynamic environmental changes such as mergers, AGN feedback, and starbursts on short timescales. 
We also trace the evolution of grain size.
Under the two-size approximation, small and large grains form and evolve through different but interrelated processes.
By combining grain size and chemical composition information, we can investigate the evolution of the intrinsic dust extinction curve and generate realistic mock observations, even at high redshifts.

Figure~\ref{fig:dtm_grain} shows the dust properties of a sample spiral galaxy at $z=0.8$.
The left panel shows the intrinsic extinction curve based on the average composition of dust inside $3R_{\rm 50,3D}$ using Mie theory and grain optical constants from~\cite{Weingartner2001}.
The total extinction curve is shown in grey, and the cyan and magenta lines represent the carbon-based and silicon-based grain species, respectively.
The contributions from small and large grains are shown with different line styles (dotted and dashed lines for small and large grains, respectively), and the grain size distribution is also presented in a sub-panel in the upper right corner.

The Milky Way (MW) average extinction curve \citep{Fitzpatrick&Massa2007}, the Small Magellanic Cloud (SMC) extinction curve \citep{Pei1992}, and the Calzetti attenuation curve for starburst galaxies \citep{Calzetti2000} are also given for reference.
We note that, unlike the MW and SMC extinction curves, the Calzetti curve represents an attenuation law that encapsulates both absorption and scattering effects in a mixed star–dust geometry (for reference, see \citealt{Narayanan2018}, which illustrates how Calzetti-like attenuation laws can emerge from different extinction curves depending on the geometry between stars and dust).
The sample \nc\ galaxy shows a UV-to-optical slope similar to the Calzetti attenuation curve, but notably with a clear 2175\r{A} bump as shown in the MW curve produced by small carbonaceous grains.

The spatial distribution of the dust-to-metal mass ratio (DTM) is also presented in the middle panel.
The mean DTM inside $3R_{\rm 50,3D}$ is $\sim 0.4$. 
We find that the DTM of the dust lanes following the spiral arms (see the RGB composite mock image in the right panel, generated using SKIRT with the SDSS $u$-, $g$-, and $r$-bands mapped to blue, green, and red, respectively) is nearly 0.5 or above, but shows various values over a wide range.
We notice a large cavity of DTM ratio in the centre and small ones in the spiral arms.
They are the sites of enhanced star formation, where SN shocks destroy a large fraction of the dust.
The central DTM cavity is of particular significance, because it seriously affects the apparent light distribution of galaxies and thus morphology determination at high redshifts \citep{2025ApJ...992...92B}.

\subsection{Cluster merger}

\begin{figure*}[!t]
    \centering
    \includegraphics[width=17cm]{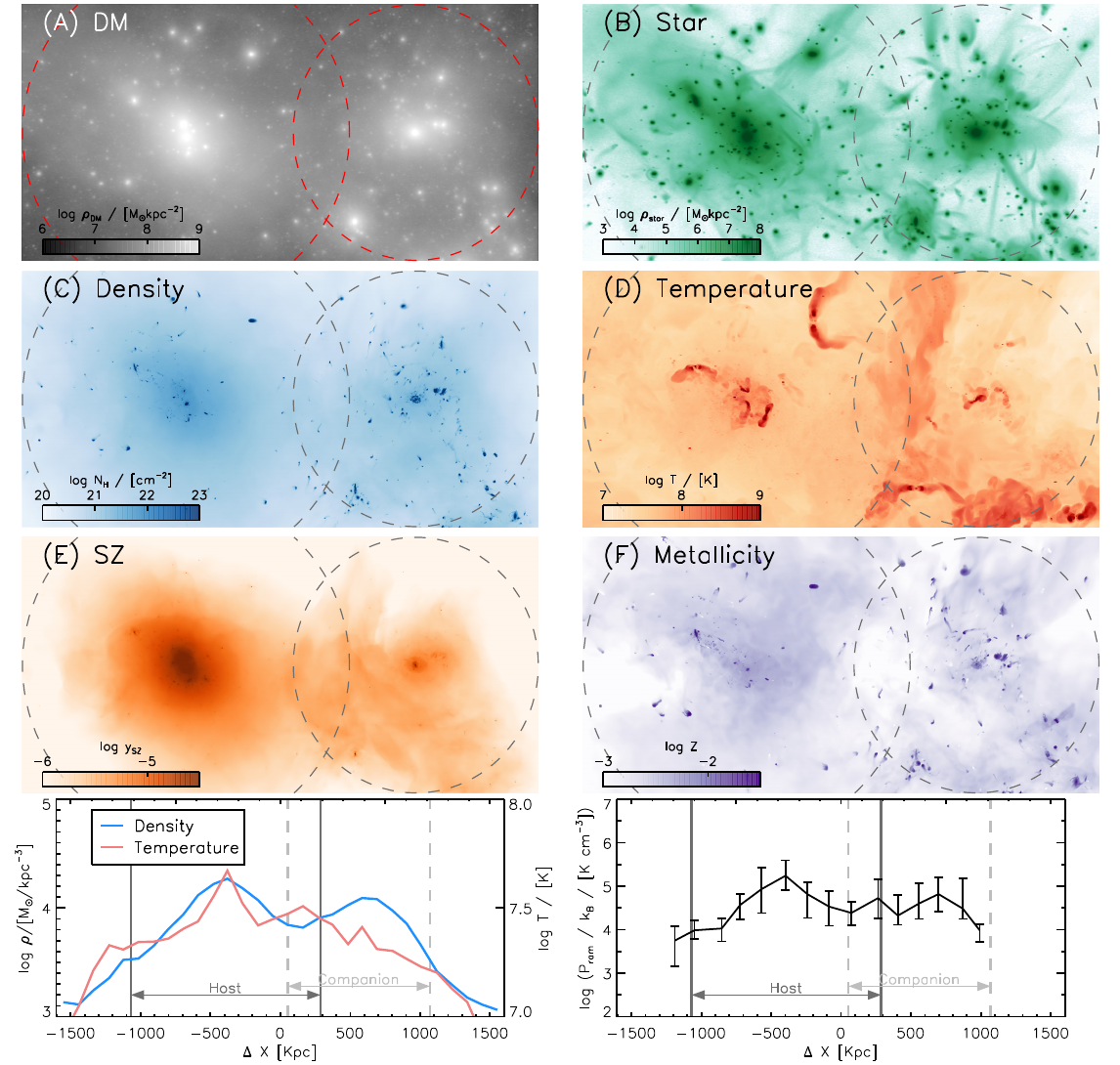}
    \caption{
    Multi-component view of the merging clusters at $z=0.73$ with different components (six upper panels) and one-dimensional profiles of the merging clusters (bottom two panels).
    From (A) to (E), DM surface density, stellar surface density, gas column density, temperature, the Sunyaev-Zeldovich signal by the $y$-parameter, and metallicity are shown, in which virial radii of the clusters are drawn with dashed circles.
    The compression of ICM caused by the major merger can be seen by a high contrast of $y$-parameter at the merger front.
    The bottom-left panel shows the one-dimensional profiles of density (blue solid line) and temperature (red solid line) along the collision axis.
    The locations of the virial radii are shown with the vertical lines.
    The bottom-right panels display the median ram pressure profiles of galaxies along the collision axis.
    Galaxies at the merger boundary ($0\,{\rm \kpc}\lesssim \Delta X \lesssim 300\, {\rm kpc}$) undergo higher external pressure compared to those on the far side of the merger front.
    }
    \label{fig:merging}
\end{figure*}

The main halo in the \nc\ simulation is approaching a major merger event (mass ratio 1:3.2).
The second-largest halo lies in close proximity, with its first encounter expected at $z \approx 0.774$ and final coalescence at $z \approx 0.35$, according to the DM-only pilot simulation.
Figure~\ref{fig:merging} illustrates a comprehensive, multi-component view of the main and secondary halos involved in the major merger.
The upper six panels display: (A) DM density, (B) stellar density, (C) gas column density, (D) gas temperature, (E) the Sunyaev-Zeldovich (SZ) signal computed using the Compton $y$-parameter, which traces the integrated pressure along the line of sight \citep[][]{Nelson24}, and (F) gas metallicity.
The two halos are projected onto a plane containing the collision axis, providing a clear view of their relative structure and separation.
Red and black dashed circles indicate the virial radii of the primary and secondary halos, respectively.

The bottom-left panel presents the one-dimensional profiles of gas density (blue solid line) and temperature (red solid line) measured along the collision axis.
The virial radii of the main and companion clusters are marked by the dark grey solid line and the light grey dashed line, respectively.
There is a noticeable increase in temperature and pressure near the interface, indicating that compressive heating is occurring in the ICM.
This feature is also evident in the Panels D and E.
In contrast, the gas metallicity in this region remains low, suggesting that the high-temperature gas is not a product of metal-enriched outflows from galaxies.
This supports the interpretation that the outer ICM of both clusters is being compressed as a direct consequence of the ongoing cluster-cluster merger.

The bottom right panel of Fig.~\ref{fig:merging} shows the median ram pressure experienced by galaxies along the collision axis.
Notably, galaxies located between the two clusters are subject to elevated ram pressure compared to galaxies at similar distances on the far side of the merger front.
This enhancement is a consequence of the ICM compression, indicating that galaxies can be affected by cluster-cluster mergers, even though they are treated as collisionless components in large-scale dynamics.
Such an increased ram pressure may lead to triggered SF or AGN activity in some galaxies \citep[][]{Owen99, Miller03, Stroe14, Stroe21}.
In this context, \nc\ provides a valuable platform for investigating the impact of large-scale mergers on galaxy evolution.

 \subsection{Quenching of galaxies in Clusters}

\begin{figure}[!t]
    \resizebox{\hsize}{!}{
    \includegraphics{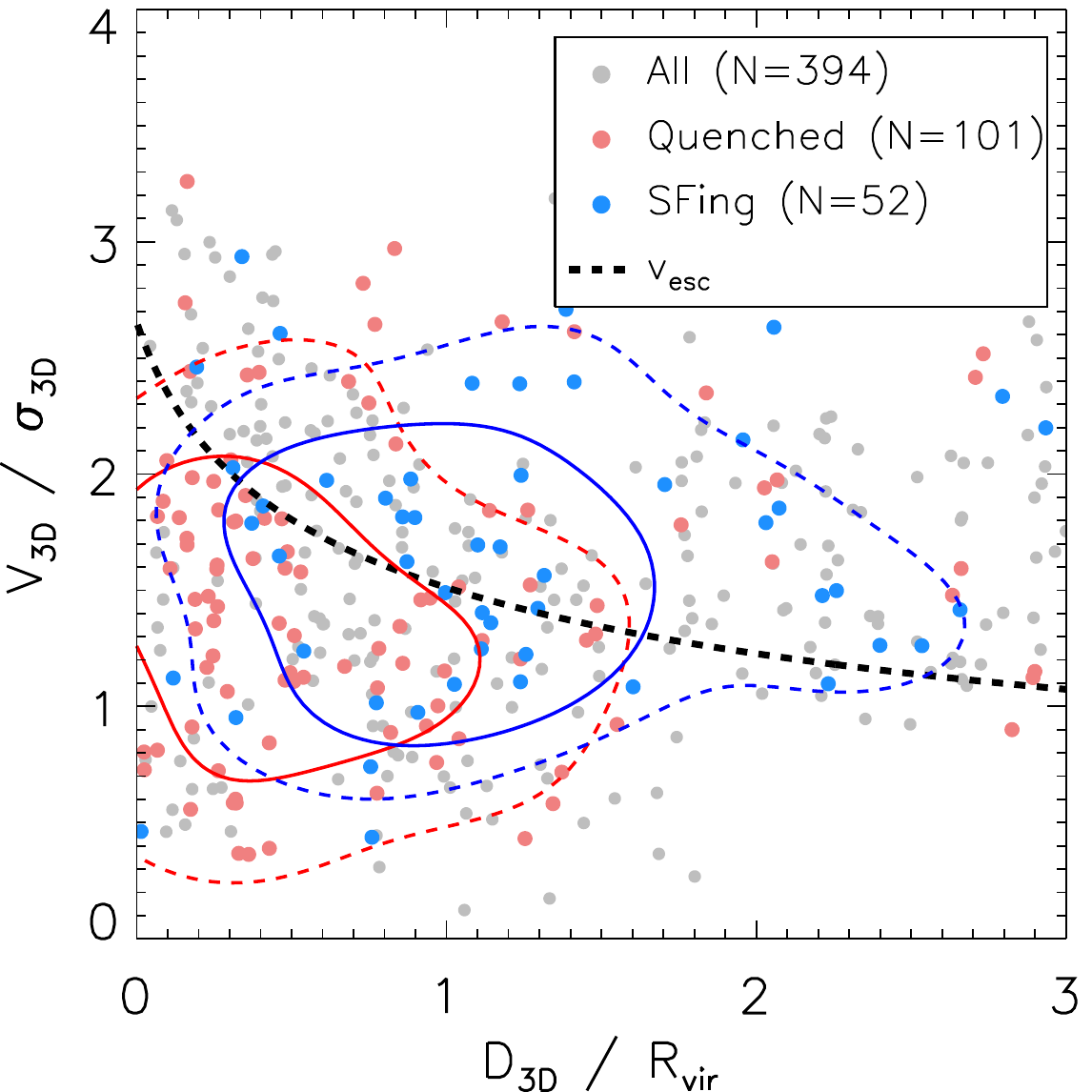}}
    \caption{
    Phase space diagram of galaxies in two dark matter halos with $M_{\rm vir} > 10^{13}\,M_{\odot}$ at $z = 1$.
    The black dashed line is the escape velocity curve profile of the more massive halos.
    Galaxies with $M_{*} > 10^{9}\,M_{\odot}$ are classified into two groups, either they are star-forming ($b > 0.1$) or quenched ($b < 0.1$).
    See the texts for the definition of the parameter $b$.
    Their $0.5\,\sigma$ and $1\,\sigma$ distributions are displayed with the contours with dashed line and solid line, respectively.
    Even at a high redshift of $z = 1$, a clear separation between the distributions of star-forming and quenched galaxies is seen.
    \label{fig:psd}}
\end{figure}

Red and dead galaxies preferentially reside in high-density environments in the local Universe \citep[][]{Dressler1980, Lewis02, Hogg04, Kauffmann04}.
In contrast, these regions likely experienced elevated star formation activity in the early Universe \citep{2022ApJ...941....5J}.
Observations suggest that substantial environmental quenching was already underway at high redshifts.
For example, \cite{Gerke07} reported an excess of blue galaxies in low-density regions (groups and the field) at $z\lesssim1$.
Furthermore, the emergence of a well-defined colour-magnitude relation in galaxy clusters at $z\sim 1$ \citep[][]{Ellis97, Kodama98, Stanford98, Mei06a, Mei06b} supports the idea of early galaxy formation in dense environments.
The presence of a red sequence on the colour-magnitude plane at these redshifts is considered a hallmark of such accelerated evolution.
Evidence of rapid environmental influence extends to even higher redshifts \citep[e.g., $z>2$;][]{Steidel05}, and JWST observations have begun tracing the formation and evolution of galaxies in overdense regions beyond $z \sim 5$ \citep[e.g.,][]{Alberts24, Morishita25}.
These findings suggest that environmental effects may have already played a significant role in shaping galaxy populations at $z>1$.

In this context, \nc\ is particularly well suited for investigating the origin of SF quenching at high redshifts, as it extends beyond $z=1$ and contains several thousand satellite galaxies residing in high-density environments.
For example, we select two halos with $M_{\rm vir} > 10^{13}\,M_{\odot}$ at $z = 1$.
Galaxies with $M_{*} > 10^{9}\,M_{\odot}$ located within three virial radii of these halos are selected and classified based on their SF activity.
Specifically, galaxies are divided into star-forming ($b>0.1$) or quenched ($b<0.1$) populations, where $b = t_{\rm Universe} \times {\rm SFR} / M_{*}$ is the birth rate parameter widely used to quantify the degree of SF quenching across a broad range of redshift and stellar masses, with $t_{\rm Univserse}$ denoting the age of the Universe \citep[][]{Franx08, Lotz19, Park22, Rhee24}.
The SFR is measured over a $100\,{\rm Myr}$ timescale.
At this stage, galaxies with $M_{*} < 10^{9}\,M_{\odot}$ are excluded from the classification, as the majority of these low-mass galaxies exhibit negligible or no SF activity.
Their rapid quenching is likely driven by their shallow gravitational potentials, which make them highly susceptible to environmental processes within clusters.

Figure~\ref{fig:psd} presents the phase-space distribution of the galaxies, with all galaxies shown in grey, and quenched and star-forming populations in red and blue, respectively.
Positions and velocities are normalised by the virial radius and velocity dispersion of the host clusters.
The black dashed curve denotes the escape velocity profile of the most massive halo, providing a reference for the dynamical boundaries of the clusters.
To illustrate the spatial concentration of each population, $0.5\sigma$ and $1\sigma$ contours are overlaid for quenched (red lines) and star-forming galaxies (blue lines).

The quenched galaxies (red circles) are primarily concentrated within the virial radius, whereas star-forming galaxies tend to occupy the outer regions, consistent with them being a recently-accreted population \citep[][]{Rhee17}.
This clear separation between the two distributions suggests a scenario in which galaxies are rapidly quenched following infall into their host clusters.
A similar separation between quenched and star-forming populations has been observed in clusters at the local Universe, where phase-space diagrams reveal strong correlations between the quenching status and the orbital history of galaxies \citep[e.g.,][]{Oman16, Rhee20, Oman21}.
These findings highlight the predictive power of phase-space analysis that extends beyond the local universe and remains effective even at high redshifts \citep[e.g.,][]{Muzzin14, Kim23}.

Given that the environmental effects in the clusters are already evident by $z=1$ \citep[e.g.,][]{Gay10}, the \nc\ simulation is well positioned to support a range of follow-up studies.
One important direction is to investigate the physical mechanisms driving rapid quenching, which likely account for the lack of star-forming galaxies in cluster core regions.
Beyond the impact on newly-accreted satellites, the simulation also offers an opportunity to examine the evolution of the Butcher-Oemler effect \citep[][]{BO78} at high redshifts, providing insights into the pace of galaxy transformation in high-density environments.
Such studies will be particularly valuable when interpreted alongside recent JWST observations tracing early galaxy populations in overdense regions.

\section{Summary}\label{sec:summary}

This paper introduces the current status of the \nc\ simulation and outlines potential scientific investigations it enables.
A cluster region is selected from a cosmological volume with a side-length of $100\,h^{-1}\Mpc$, and is subsequently re-simulated at high resolutions, incorporating a comprehensive and modern astrophysics model (see Sect.~\ref{sec:method}).
In addition to the standard prescriptions for SF and feedback models from stars and super-massive black holes (SMBHs), \nc\ performs on-the-fly modeling of dust evolution and chemical enrichment for ten species (H, D, He, C, N, O, Mg, Fe, Si, and S).
These capabilities allow for the investigation of previously inaccessible aspects of galaxy evolution in a cosmological context.
For instance, shaping galaxy morphologies at high redshifts, informed by dust extinction from realistic dust distributions within galaxies has been studied in \cite{2025ApJ...992...92B}.
Figure~\ref{fig:mock} highlights these novel features, presenting mock images of \nc\ galaxies that closely resemble real observations.

Because it simulates one of the most complex regions of the Universe at very high resolution, \nc\ requires computationally intensive calculations.
For example, the simulation reached $z=0.8$, spending more than $10^8$ CPU hours. 
These extensive calculations have been made possible thanks to significant scalability enhancements in the RAMSES code, achieved through hybrid parallelization with OpenMP and MPI \citep[][]{yomp}.
The updated RAMSES code shows efficient performance in using numerous CPU cores simultaneously, with which the \nc\ simulation is running using 7,680--15,360 threads, depending on the speed and resource situations.

By achieving high resolution (e.g., $68\,{\rm pc}$ for spatial resolution) in a cluster simulation, \nc\ enables detailed studies of galaxy evolution across a wide mass range within cosmological cluster environments.
As a result, several key scientific investigations become feasible.
The excellent stellar mass resolution of \nc\ ($\sim 2\times10^{4}\,M_{\odot}$) allows detection of LSB features down to $\mu \sim 33.5\,{\rm mag}\,{\rm arcsec}^{-2}$ at $z \sim 0.8$.
Importantly, the cluster environment targeted by \nc\ is particularly rich in LSB structures, including tidal debris, UDGs, and ICL.
Figure~\ref{fig:LSB} showcases a variety of such LSB features captured in the simulation.

One of the key features of \nc\ is an embedding of the Monte-Carlo tracer particle to track the Lagrangian motion of gas cells.
With the successful capture of the small-scale galactic features via recent hydrodynamical simulations, we now want to investigate the detailed trajectory and evolutionary pathway of gas to build those features.
The tracer particles in \nc\ recover the true distribution of gas. 
Figure~\ref{fig:tracer} shows an example of the prospect of tracer particles.
\nc\ with tracer particles would extend our interests to various topics, such as gas recycling in star formation or MBH accretion, distinguishing the inflow and outflow gases.

Although \nc\ does not yet extend to the local Universe, it reaches the formation epoch of the cluster and provides critical insights into the evolution of galaxies in the first half of the history of the Universe.
As shown in Fig.~\ref{fig:rps}, ram pressure is already affecting satellite galaxies at this epoch, with many displaying jellyfish-like morphologies.
Galaxy D in Fig.~\ref{fig:rps}, for example, exhibits complete gas removal.
These early interactions contribute to the emergence of gas-deficient galaxies well before the present day.
Moreover, likely due in part to the effects of ram pressure, Figure~\ref{fig:psd} reveals a clear separation between star-forming and quenched galaxies in the early clusters in the phase space, suggesting that environmental quenching is already effective at $z \sim 1$.
The presence of passive galaxies in such early structures raises important questions about the onset and efficiency of environmental effects in the high-redshift Universe.

Advancing toward a better understanding of dust evolution through cosmic time might be one of the most important and urgent tasks in the JWST era.
Because not only the chemical composition of grain species but also their geometric distribution matters, a physically motivated theoretical prediction in a cosmological context is crucial.
Although many cosmological simulations have included dust evolution on-the-fly \citep{2018MNRAS.479.2588G,2019MNRAS.485.1727H,2021A&A...653A.154T,Granato21,2023MNRAS.519.5987L}, \nc\ is one of the pioneering cosmological simulations to simultaneously track both the grain size distribution and chemical composition of dust down to redshifts $z<1$.

Figure~\ref{fig:dtm_grain} clearly shows how calculations with different grain species (carbon- and silicon-based) and different size bins successfully reproduce the extinction curve.
Despite the fact that our dust model is built on a relatively simple prescription compared to the real universe, we hope that our results can serve as a cornerstone for a better understanding of dust evolution from the high-redshift universe to the local universe.
\nc\ represents a first step toward that goal.

\nc\ at $z\sim0.8$ captures a notable pre-merger configuration, in which the main cluster and the second most massive halo are approaching one another.
Figure~\ref{fig:merging} illustrates a multi-dimensional view of the two halos at this early merger stage.
At the interface between the halos, signatures of gas compression are evident, particularly at the merger front, wherein galaxies are subject to enhanced ram pressure.
These conditions suggest that cluster-cluster mergers can significantly influence galaxy evolution, even prior to the core passage.
The results highlight the possibility that such mergers may trigger SF in some galaxies while accelerating quenching in others.
\nc\ thus offers a powerful platform for studying how large-scale dynamical events reshape galaxy populations during the assembly of massive structures.
\nc\ will soon reveal the complete merger between the clusters at around $z \approx 0.5$. This will allow a number of exciting investigations on galaxy evolution as well as cluster evolution. 

The \nc\ simulation enables a broad range of follow-up studies, spanning scales from SMBH to the cosmic web.
A list of ongoing science cases is available on the simulation website.\footnote{\href{https://gemsimulation.com}{https://gemsimulation.com}}
To support and encourage further research, we provide public access to the \nc\ dataset via the website, including the galaxy and halo catalog, subset of raw simulation outputs, and mock images.
We welcome participation from the community in expanding the scientific exploration enabled by this simulation.

A critical shortcoming of \nc\ is that it is only one cluster. Therefore, it cannot be representative.
However, it is already too challenging to get one cluster done for such resolutions.
After 3 years of nearly-continuous runs, we only reached $z=0.8$\footnote{As we speak, \nc\ is passing $z=0.70$.}, and we expect to require a similar or greater amount of computing time to reach the local universe.
Now that we see the beauty and power of high-resolution data, we want to continue the march to simulate more clusters.
But for that to be possible, a dramatically improved numerical approach would be necessary.

\begin{acknowledgements}

S.K.Y. acknowledges support from the Korean National Research Foundation (RS-2025-00514475 and RS-2022-NR070872). 

This work was granted access to the HPC resources of KISTI under the allocations KSC-2021-CRE-0486, KSC-2022-CRE-0088, KSC-2022-CRE-0344, KSC-2022-CRE-0409, KSC-2023-CRE-0343, KSC-2024-CHA-0009, and KSC-2025-CRE-0031 and of GENCI under the allocation A0150414625 and A0180416216.
The large data transfer was supported by KREONET which is managed and operated by KISTI. 
J.R was supported by the KASI-Yonsei Postdoctoral Fellowship and was supported by the Korea Astronomy and Space Science Institute under the R\&D programme (Project No. 2025-1-831-02), supervised by the Korea AeroSpace Administration.
This work was partially supported by the Institut de Physique des deux infinis of Sorbonne Université and by the ANR grant ANR-19-CE31-0017 of the French Agence Nationale de la Recherche.
J.K. is supported by KIAS Individual Grants (KG039603) at the Korea Institute for Advanced Study.
T.K. is supported by the National Research Foundation of Korea (RS-2025-00516961) and the Yonsei Fellowship funded by Lee Youn Jae.

\end{acknowledgements}.
\bibliographystyle{aa}
\bibliography{references}

\end{document}